\documentclass[journal]{IEEEtran}
\usepackage{cite}
\usepackage{mathtools}
\usepackage{gensymb}
\usepackage[caption=false,font=footnotesize]{subfig}
\usepackage[ruled,vlined]{algorithm2e}
\usepackage{amsmath,amssymb,amsfonts,amsthm}
\usepackage{pifont}
\usepackage{textcomp}
\usepackage{tikz}
\usepackage{tikzscale}
\usepackage{circuitikz}
\usepackage[acronyms,nonumberlist]{glossaries}

\newacronym{AI}{AI}{artificial intelligence}
\newacronym{BS}{BS}{base station}
\newacronym[plural=UEs,
            longplural={user equipment}]{UE}{UE}{user equipment}
\newacronym{UL}{UL}{uplink}
\newacronym{DL}{DL}{downlink}
\newacronym{PASS}{PASS}{pinching-antenna system}
\newacronym{PA}{PA}{pinching antenna}
\newacronym{LoS}{LoS}{line-of-sight}
\newacronym{NLoS}{NLoS}{non-line-of-sight}
\newacronym{AWGN}{AWGN}{additive white Gaussian noise}
\newacronym{SNR}{SNR}{signal-to-noise ratio}
\newacronym{SINR}{SINR}{signal-to-interference-plus-noise ratio}
\newacronym{SIR}{SIR}{signal-to-interference ratio}
\newacronym{CDF}{CDF}{cumulative distribution function}
\newacronym{PDF}{PDF}{probability density function}
\newacronym{RV}{RV}{random variable}
\newacronym{BM}{BM}{Brownian motion}
\newacronym{TA}{TA}{tangent approximation}
\newacronym{REM}{REM}{radio environment map}
\newacronym{NN}{NN}{nearest neighbor}
\newacronym{BI}{BI}{bilinear interpolation}
\newacronym{OMA}{OMA}{orthogonal multiple access}
\newacronym{NOMA}{NOMA}{non-orthogonal multiple access}
\newacronym{LSF}{LSF}{large-scale fading}
\newacronym{SSF}{SSF}{small-scale fading}
\newacronym{RIS}{RIS}{reconfigurable intelligent surface}
\newacronym{FAS}{FAS}{fluid antenna system}
\newacronym{SU}{SU}{single-user}
\newacronym{BER}{BER}{bit error rate}
\newacronym{iid}{i.i.d.}{independent and identically distributed}
\newacronym{CSI}{CSI}{channel state information}
\newacronym{WG}{WG}{waveguide}
\newacronym{OP}{OP}{outage probability}
\newacronym{SWIPT}{SWIPT}{simultaneous wireless information and power transfer}
\newacronym{TS}{TS}{two-stage}
\newacronym{SCP}{SCP}{scattered channel power}
\newacronym{PLE}{PLE}{path loss exponent}
\newcommand{\tick}{\ding{51}}
\newcommand{\cross}{\ding{55}}

\newtheorem{lemma}{Lemma}
\newtheorem{rem}{Remark}

\usepackage[compact]{titlesec} %for squeezing around titles etc
\titlespacing\section{0pt}{3pt plus 2pt minus 1pt}{3pt plus 2pt minus 1pt}
\titlespacing\subsection{0pt}{3pt plus 2pt minus 1pt}{2pt plus 1pt minus 2pt}
\titlespacing\subsubsection{0pt}{6pt plus 4pt minus 1pt}{0pt plus 2pt minus 1pt}

\begin{document}

\title{How Robust are Pinching-Antenna Systems to Practical Deployment Imperfections?}
\author{Amy~S.~Inwood,~\IEEEmembership{Member,~IEEE,} Peter~J.~Smith,~\IEEEmembership{Fellow,~IEEE,}
Hyundong~Shin,~\IEEEmembership{Fellow,~IEEE,} and~Michail~Matthaiou,~\IEEEmembership{Fellow,~IEEE\vspace{-0.75cm}}

%\thanks{This work was supported by the U.K. Engineering and Physical Sciences Research Council (EPSRC) grant (EP/X04047X/2) for TITAN Telecoms Hub. The work of P. J. Smith was supported by the Marsden Fund Council from New Zealand Government funding, managed by Royal Society Te Apārangi. The work of H. Shin was supported by the National Research Foundation of Korea (NRF) grant funded by the Korean government (MSIT) under RS-2025-00556064 and by the MSIT (Ministry of Science and ICT), Korea, under the ITRC (Information Technology Research Center) support program (IITP-2025-RS-2021-II212046) supervised by the IITP (Institute for Information \& Communications Technology Planning \& Evaluation). The work of M. Matthaiou was supported by the European Research Council (ERC) under the European Union’s Horizon 2020 Research and Innovation Programme (grant agreement No. 101001331).}

\thanks{A. S. Inwood is with the Centre for Wireless Innovation (CWI), Queen’s University Belfast, Belfast BT3 9DT, U.K. (e-mail: a.inwood@qub.ac.uk).}

\thanks{P. J. Smith is with the School of Mathematics and Statistics, Victoria University of Wellington, Wellington, New Zealand (e-mail: peter.smith@vuw.ac.nz).}

\thanks{H. Shin is with the Department of Electronics and Information Convergence Engineering, Kyung Hee University, Yongin-si, Gyeonggi-do 17104, Republic of Korea (e-mail: hshin@khu.ac.kr).}

\thanks{M. Matthaiou is with the Centre for Wireless Innovation (CWI), Queen’s University Belfast, BT3 9DT Belfast, U.K., and also with the
Department of Electronic Engineering, Kyung Hee University, Yongin-si, Gyeonggi-do 17104, Republic of Korea (e-mail: m.matthaiou@qub.ac.uk).}
}
 
\maketitle

\begin{abstract} 
This paper investigates the robustness of pinching-antenna systems (PASSs) to three practical deployment imperfections: pinching-antenna (PA) positioning errors, imperfect channel state information (CSI), and line-of-sight (LoS) blockages. We develop analytical frameworks to quantify the performance degradation caused by each imperfection. For PA positioning errors, we employ a Brownian meander model to characterize the distribution and mean of the resulting signal-to-noise ratio (SNR) loss. The results show that positioning errors can diminish the gains from long-term position selection, with the loss becoming more pronounced under stronger shadow fading. For imperfect CSI, we characterize the distribution and moments of the true scattered channel power at the position selected using noisy channel estimates. The resulting selection penalty is governed by the ratio of noise variance to the number of channel measurements, while the upper tail of the selected channel power retains the same behavior as under perfect CSI. Finally, we develop a spatially consistent blockage model and derive the distribution of the optimum long-term SNR for arbitrary blockage configurations, together with a numerically tractable expression for a single blockage. Simulation results validate the analyses and highlight a fundamental property of PASSs: although positioning errors and imperfect CSI can reduce the gains from position selection, good performance can still be achieved provided that an LoS link is available. In particular, even when a large proportion of the waveguide is blocked, PA repositioning can establish LoS links and enable performance close to that of a fully unblocked system for the considered configurations.
\end{abstract}

\begin{IEEEkeywords}
Pinching-antenna systems, Rayleigh fading, Ricean fading, statistical analysis.
\end{IEEEkeywords}

\section{Introduction}

%As the number of connected devices and the demand for data-intensive applications continue to grow, so do the demands placed on mobile network capacity. Moving to higher carrier frequencies can meet this demand, but this also degrades propagation conditions through increased path loss, attenuation, and shadowing.

Flexible antenna technologies \cite{ding_flexible_2025} have attracted growing interest since they allow the radio propagation environment itself to be controlled. \Glspl{FAS} can reconfigure on the wavelength scale and improve \gls{SSF} performance, but they cannot establish new \gls{LoS} paths. In contrast, \glspl{RIS} can create \gls{LoS} links, but doing so introduces an additional propagation hop, compounding path losses.

\Glspl{PASS}, first proposed by NTT DOCOMO \cite{fukuda_pinching_2022}, avoid this trade-off. A small dielectric element, known as a \gls{PA}, is attached to a dielectric \gls{WG} fed by a \gls{BS}. The signal radiates from where the \gls{PA} is placed, allowing the establishment of an \gls{LoS} link without an additional hop. Also, \glspl{PASS} are expected to be constructed with simple and inexpensive passive components, making them cheaper than an extra \gls{BS}.

Existing studies of \glspl{PASS} have largely concentrated on \gls{PA} position selection or investigating performance under ideal conditions. Practical deployments, however, will be subject to imperfections, including \gls{PA} positioning errors, imperfect \gls{CSI}, and \gls{LoS} blockages, each of which can affect the gains that \glspl{PASS} promise. Quantifying these effects is therefore essential to understanding the true benefits of the technology and to designing systems that remain robust in realistic environments.

\subsection{Review of Related Literature}

As an emerging technology, there has been significant recent research attention on \glspl{PASS}. However, most of this work has focused on ideal systems. In this section, existing works on \gls{PASS} imperfections, namely on \gls{PA} position errors, imperfect \gls{CSI} and \gls{LoS} blockages, are detailed and compared.

A focus of \gls{PASS} study has been the determination of optimal \gls{PA} positions on the \gls{WG}. The optimization of various metrics, such as \gls{SNR}, sum rate, \gls{BER} and \gls{OP}, via \gls{PA} position selection has been studied for a range of systems \cite{alaaeldin_ber_2026,qian_pinching_2026, ding_flexible_2025,wang_antenna_2025,tyrovolas_performance_2026}. Despite this, there has been limited analysis of the impact of a \gls{PA} not being precisely located at the optimal point. This could happen for a variety of reasons, such as discrete port locations or mechanical tolerances. In \cite{inwood_analysis_2026}, a general random error term was added to the selected position to represent a range of errors, including position error and outdated \gls{CSI}. However, this did not explicitly model \gls{PA} displacement and was only investigated via simulation, making the analysis of \gls{PA} position errors a key open research problem. 

Additionally, there have been limited analytical studies of the impact of imperfect \gls{CSI} on \glspl{PASS}. Most work on \gls{PASS} \gls{CSI} has focused on acquisition methods \cite{zhou_channel_2025,xiao_channel_2025,guo_transformer_2026} rather than performance analysis. In \cite{zhou_channel_2025}, a novel channel estimation method was proposed for a millimeter wave \gls{PASS}, where subarrays of \glspl{PA} allowed the effective emulation of a larger aperture array and enabled accurate estimation of multipath propagation parameters. The works in \cite{xiao_channel_2025} and \cite{guo_transformer_2026} both proposed \gls{AI}-based approaches to \gls{CSI} acquisition in \glspl{PASS} with a single \gls{WG} and multiple \glspl{PA}. Channel estimation in static scenarios under both \gls{LoS} and \gls{NLoS} propagation was addressed in \cite{xiao_channel_2025}, while \cite{guo_transformer_2026} considered the prediction of pure \gls{LoS} links in highly mobile scenarios. As novel methods were the focus of these works, limited performance analysis was undertaken.

A numerical study of \gls{CSI} errors in \gls{LoS} channels in a physical layer security setting was conducted in \cite{cenzig_physical_2026}. For a purely \gls{LoS} \gls{PASS}, the effects of \gls{CSI} error and quantized feedback on the secrecy rate were simulated, and it was found that the degradation of the \gls{PASS} was less than a comparable fixed antenna system. The work in \cite{modi_mobility_2026} conducted an analytical study on \gls{CSI} aging caused by the Doppler effect in highly mobile scenarios. Assuming a purely scattered channel, a conditional outage probability was derived. Therefore, analytical studies of imperfect channel estimation in \glspl{PASS} are limited.

A number of studies have considered blockage in \glspl{PASS}, however the majority have not conducted analytical performance analyses or considered simplified systems. One method of modeling blockages is through stochastic geometry, where the probability of a blockage is determined by a \gls{RV}. While this simplifies analysis, it does not typically allow for specific environmental configurations to be analyzed. The works in \cite{ding_los_2025,wang_outage_2026,hanif_multicasting_2026,jiang_spatially_2026} all considered systems where an \gls{LoS} link was controlled by a distance-dependent \gls{RV}, and if this \gls{LoS} link was blocked, service was rendered impossible. For a single-\gls{UE}, single \gls{PA} system, \cite{ding_los_2025} and \cite{wang_outage_2026} derived expressions for the \gls{OP}, while \cite{ding_los_2025} also investigated the ergodic rate of a system where multiple \glspl{UE} were served by a dedicated \gls{WG} with a single \gls{PA}. In \cite{hanif_multicasting_2026}, the \gls{PA} positions were optimized to maximize the minimum \gls{SNR} in a multicast scenario, and in \cite{jiang_spatially_2026}, \gls{PA} positions were optimized to maximize the average \gls{SNR} subject to minimum energy harvesting requirements in an \gls{SWIPT} system.

In \cite{sun_stochastic_2026}, the channel between each \gls{PA} and \gls{UE} was modeled as either \gls{LoS} or \gls{NLoS}, determined using a rectangular Boolean modeling method. For this scenario, an expression for the \gls{OP} was derived. The work in \cite{xu_pinching_2026} derived the average \gls{SNR} for a system where an \gls{NLoS} channel component modeled by scattering clusters was always present, and the existence of an \gls{LoS} component was determined by a Bernoulli \gls{RV}. While this is a more realistic channel configuration, shadow fading and correlation across the \gls{WG} were not modeled.

A more spatially consistent approach is to deterministically model blockages. This flexible approach allows specific realistic scenarios to be analyzed through the specification of blockage locations. In \cite{zhao_robust_2026}, a \gls{PASS} system with deterministic blockages was considered, where the channel was either \gls{LoS} or fully blocked. This work focused on physical layer security under imperfect \gls{CSI} and proposed a joint optimization framework where the information and additive noise beamforming, the \gls{PA} positions, and the individual \gls{PA} power-ratio allocation were optimized to maximize the sum rate. 

The studies in \cite{wang_on_2026, wang_pinching_2025,xie_pinching_2026} all considered systems with \gls{NLoS} components and an \gls{LoS} component whose presence was determined by the location of known blockages. In \cite{wang_on_2026}, the \gls{OP} and ergodic rate were derived under both Ricean and Rayleigh \gls{SSF}. However, the large scale fading was purely deterministic, as shadow fading was not modeled. The sum rate was maximized through the joint optimization of \gls{WG} assignment and \gls{PA} placement in \cite{wang_pinching_2025} for a system with a general number of \glspl{WG}, \glspl{PA} and \glspl{UE}. In this work, both the \gls{LoS} and \gls{NLoS} components were deterministic, requiring highly detailed knowledge of the environment. The sum rate was also optimized in \cite{xie_pinching_2026} for a multi-\gls{UE} and multi-\gls{WG} system. Here, the \gls{LoS} component was deterministic while the \gls{NLoS} component was independent Rayleigh fading. 

Therefore, it can be seen that while existing works have considered \glspl{PASS} with blockages, these have either focused on optimization or have made simplifying assumptions around the geometry or channel configuration to enable analysis. No existing works have conducted performance analyses of \glspl{PASS} involving Ricean channels with correlated shadow fading and blockages, and this remains an open research problem.

\subsection{Contributions}

Position selection and performance analysis for \glspl{PASS} are typically carried out under ideal conditions, but the gains of a \gls{PASS} depend on a number of assumptions holding in practice. These include that the \gls{PA} can be placed exactly at the intended position, that the channel used to select that position is accurately known, and that the \gls{LoS} link is unobstructed. In reality, the \gls{PA} may not sit exactly at the chosen point because of discrete port locations, mechanical tolerances or measurement errors, and channel estimates can be corrupted by noise and other errors. Also, with \glspl{WG} spanning meters, some positions will be blocked by obstacles, so a strong \gls{LoS} link cannot inherently be assumed to hold across the full \gls{WG}.

Understanding the effects of imperfections is crucial as \glspl{PASS} derive their advantage from placing a \gls{PA} precisely where the channel is strongest. Under correlated shadow fading, small displacements and blocked regions change the \gls{SNR} in a structured, spatially dependent way, so their effect cannot be captured by treating them as independent noise. Without quantifying them, performance gains may be overstated and system designs cannot be made robust. However, no existing study has examined \gls{PA} positioning errors, imperfect \gls{CSI}, and \gls{LoS} blockages under channels that combine an \gls{LoS} component with a scattered component subject to correlated fading.

Motivated by this gap, we analyze the performance loss caused by these three imperfections in a \gls{PASS} operating under Ricean fading with spatially correlated shadow fading. This general framework accommodates arbitrary fading severity and provides insights into the transition from ideal scenarios to those incorporating realistic imperfections.

We make the following contributions:
\begin{itemize}
    \item Using a Brownian meander to model the long-term \gls{SNR} process on either side of the optimum, we derive accurate closed-form approximations for the density, distribution, and mean of the \gls{SNR} loss due to \gls{PA} positioning errors.
    \item We derive the mean, second moment, and variance of the true \gls{SSF} \gls{SCP} at the best \gls{PA} position selected from noisy channel estimates, together with its full distribution. These give the mean power loss in closed form, separate the effects of selection and estimation quality through a single parameter, and show that the upper tail matches the perfect-\gls{CSI} case.
    \item We propose a spatially consistent model of the long-term \gls{SNR} under an arbitrary number of blockages. The \gls{LSF} process is modeled by two independent Brownian motion processes, one for \gls{LoS} regions and one for \gls{NLoS} regions, and the observed process switches between them as the \gls{PA} crosses blockage boundaries. We derive a general expression for the distribution of the optimum \gls{SNR} and reduce it to a numerically tractable form for a single blockage.
\end{itemize}
Analytical results are verified by simulation, and the effects of key system parameters, such as \gls{WG} length and shadow fading standard deviation, are investigated. These contributions represent a cohesive treatment of the performance of a \gls{PASS} under realistic imperfections. Table \ref{tab:comparison} compares this work with the state-of-the-art. 
\begin{table*}[t]
  \caption{The contributions of this work compared to existing literature on Imperfections in PASSs.}
    \centering
    \setlength{\tabcolsep}{5pt}
    \begin{tabular}{|c|c|c|c|c|c|c|c|c|c|c|c|c|c|c|c|c|c|}
    \hline
        \textbf{Contributions} & 
        \textbf{This Work} & 
        \cite{inwood_analysis_2026} &  \cite{zhou_channel_2025} &  
        \cite{xiao_channel_2025} &  \cite{guo_transformer_2026}  &  \cite{cenzig_physical_2026} & \cite{modi_mobility_2026} & 
        \cite{ding_los_2025} & 
        \cite{wang_outage_2026} &
        \cite{hanif_multicasting_2026} &
        \cite{jiang_spatially_2026} &
        \cite{sun_stochastic_2026} &
        \cite{xu_pinching_2026} &
        \cite{zhao_robust_2026} &
        \cite{wang_on_2026} &
        \cite{wang_pinching_2025} &
        \cite{xie_pinching_2026} \\
        \hline
        Pinching antenna systems & \tick & \tick & \tick & \tick & \tick & \tick & \tick & \tick & \tick & \tick & \tick & \tick & \tick & \tick & \tick & \tick & \tick \\ \hline
        Scattered channels & \tick & \tick & \tick & \tick & \cross & \cross & \tick & \cross & \cross & \cross & \cross & \tick & \tick & \tick & \tick & \tick & \tick \\ \hline
        Correlated shadowing & \tick & \tick & \cross & \cross & \cross & \cross & \cross & \cross & \cross & \cross & \cross & \cross & \cross & \cross & \cross & \cross & \cross \\ \hline
        PA positioning errors & \tick & \tick & \cross & \cross & \cross & \cross & \cross & \cross & \cross & \cross & \cross & \cross & \cross & \cross & \cross & \cross & \cross \\ \hline
        Imperfect CSI & \tick & \cross & \tick & \tick & \tick & \tick & \tick & \cross & \cross & \cross & \cross & \cross & \cross & \tick & \cross & \cross & \cross \\ \hline
        Blockage analysis & \tick & \cross & \cross & \cross & \cross & \cross & \cross & \tick & \tick & \tick & \tick & \tick & \tick & \tick & \tick & \tick & \tick \\
        \hline
    \end{tabular}
    \label{tab:comparison}
\end{table*}

\textit{Notation}: The set of complex numbers is denoted by $\mathbb{C}$; $\Re(\cdot)$ is the real component of a complex number; $\mathbb{E}[\cdot]$ and $\mathrm{Var}[\cdot]$ are expectation and variance;  $\mathcal{N}(\mu,\sigma^2)$ and $\mathcal{CN}(\mu,\sigma^2)$ denote the Gaussian and complex Gaussian distributions with mean $\mu$ and variance $\sigma^2$; $f_X(x)$ and $F_X(x)$ are the \gls{PDF} and \gls{CDF} of random variable $X$; $\varphi(\cdot)$ and $\Phi(\cdot)$ are the standard normal \gls{PDF} and \gls{CDF}; $N(\mathbf{x};\boldsymbol{\mu},\boldsymbol{\Sigma})$ and $\Phi_n(\mathbf{x};\boldsymbol{\mu},\boldsymbol{\Sigma})$ are the multivariate normal \gls{PDF} and \gls{CDF} of dimension $n$, $\mathrm{Exp}(\cdot)$ is the exponential distribution; $P(A)$ denotes the probability of event $A$; $\mathrm{erfc}(\cdot)$ is the complementary error function;  $I_0(\cdot)$ is the zeroth-order modified Bessel function of the first kind; $f'(x)=\frac{d}{dx}[f(x)]$ and $f''(x)=\frac{d^2}{dx^2}[f(x)]$ are the first and second derivatives of $f(x)$ with respect to $x$; $\|\cdot\|_2$ denotes Euclidean distance, and $\sup_{a<x<b} (\cdot)$ and $\max_{a<x<b}(\cdot)$ denote the supremum and maximum over $x \in (a,b)$.

\section{System Model}
\label{sec:sys_mod}
\begin{figure}[ht]
    \centering
    \includegraphics[width=0.8\linewidth]{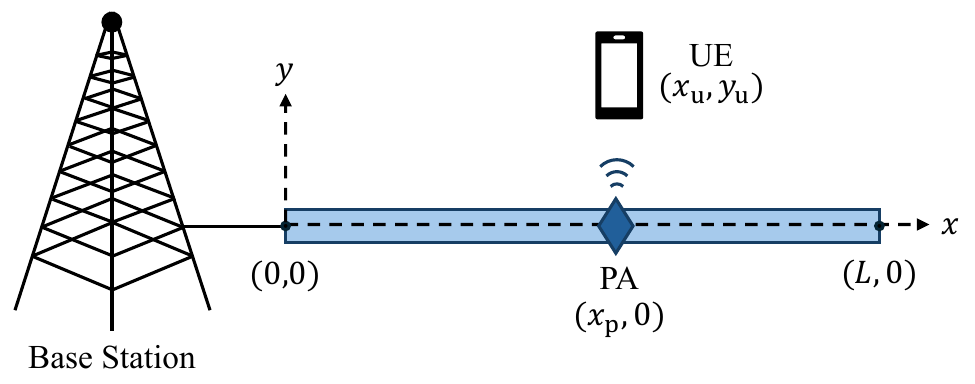}
    \caption{The studied \gls{SU} \gls{PASS}.}
    \label{fig:sys_mod} 
\end{figure}

Consider the \gls{PASS} shown in Fig.~\ref{fig:sys_mod}, 
where a single-antenna \gls{UE} at location $\boldsymbol{\psi}_\mathrm{u}=(x_\mathrm{u},y_\mathrm{u})$ is served by a \gls{BS} via a single \gls{PA}. The \gls{PA} can be located anywhere on the \gls{WG} of length $L$. It is assumed that the \gls{WG} runs along the $x$-axis, such that the \gls{PA} is located at $\boldsymbol{\psi}_\mathrm{p}(x_\mathrm{p})=(x_\mathrm{p},0)$, $0\leq x_\mathrm{p}\leq L$. \vspace{-1em}

Let the channel from the \gls{UE} to the \gls{PA} be denoted $h_\mathrm{p}(x_\mathrm{p})~\in~\mathbb{C}$. Assuming Ricean fading,
\begin{align}
    h_\mathrm{p}(x_\mathrm{p})\!&=\!\!\sqrt{\!\beta_\mathrm{LS}(x_\mathrm{p})}\Big(\!\sqrt{\!\tfrac{K}{K+1}}h_\mathrm{LoS}(x_\mathrm{p})\!+\!\sqrt{\!\tfrac{1}{K+1}}h_\mathrm{SC}(x_\mathrm{p})\!\Big)\!, \notag \\
    &=\!\!\sqrt{\!\beta_\mathrm{LS}(x_\mathrm{p})}\tilde{h}_\mathrm{p}(x_\mathrm{p}), \label{eq:htilde}
\end{align}
where $\beta_\mathrm{LS}(x_\mathrm{p})$ is the \gls{LSF} loss between the \gls{UE} and \gls{PA}, $K$ is the Ricean $K$-factor and $h_\mathrm{LoS}(x_\mathrm{p})$ and $h_\mathrm{SC}(x_\mathrm{p})$ are the \gls{LoS} and scattered channel components, respectively. The \gls{LoS} channel is modeled by the spherical wave channel model \cite{ding_flexible_2025}:
\begin{equation}  \label{sphr_ch_mod}
    h_\mathrm{LoS}(x_\mathrm{p}) = \mathrm{e}^{-j\frac{2\pi}{\lambda}d(x_\mathrm{p})},
\end{equation}
where $d(x_\mathrm{p}) = \left\|\boldsymbol{\psi}_\mathrm{u}-\boldsymbol{\psi}_\mathrm{p}(x_\mathrm{p})\right\|_2$ is the Euclidean distance between the \gls{UE} and \gls{PA}, $\lambda=\frac{c}{f_\mathrm{c}}$ is the wavelength, $c=3\times10^8$ m/s is the speed of light in a vacuum, and $f_\mathrm{c}$ is the carrier frequency. The scattered channel is modeled by Rayleigh fading, where $h_\mathrm{SC}(x_\mathrm{p})\sim \mathcal{CN}(0,1)$. Correlation is assumed to occur between the scattered channels at \gls{PA} positions $\boldsymbol{\psi}_\mathrm{p}(x_\mathrm{p})$ and $\boldsymbol{\psi}_\mathrm{p}(x_\mathrm{p}+\tau)$ only if $\tau<d_\mathrm{SS}$, where $d_\mathrm{SS}$ is the \gls{SSF} decorrelation distance. The classic path loss and shadowing model in \cite{goldsmith_wireless_2005} leads to an \gls{LSF} loss of
\begin{equation}
    \beta_\mathrm{LS}(x_\mathrm{p}) = A\,d(x_\mathrm{p})^{-\gamma}L(x_\mathrm{p}),
\end{equation}
where $A=\big(\frac{\lambda}{4\pi d_0}\big)^2$, $d_0=1$ m is the reference distance, $\gamma$ is the \gls{PLE}, $L(x_\mathrm{p})=10^{\sigma_\mathrm{SF}X(x_\mathrm{p})/10}$ is spatially correlated shadow fading driven by the correlated Gaussian process $X(x_\mathrm{p})\sim\mathcal{N}(0,1)$, and $\sigma_\mathrm{SF}$ is the shadow fading standard deviation. Assuming exponentially correlated shadow fading \cite{gudmundson_correlation_1991}, $\rho_\mathrm{LS}(\tau)=\mathbb{E}[X(x_\mathrm{p})X(x_\mathrm{p}+\tau)]=\mathrm{e}^{-\alpha|\tau|}$, where $\alpha{=}1/d_\mathrm{LS}$ and $d_\mathrm{LS}$ is the \gls{LSF} decorrelation distance. The loss within the \gls{WG} is modeled as \cite{ding_flexible_2025}
\begin{equation}
    \beta_\mathrm{W}(x_\mathrm{p}) = 10^{-\frac{\kappa}{20}x_\mathrm{p}}\mathrm{e}^{-j\frac{2\pi}{\lambda_\mathrm{g}}x_\mathrm{p}},
\end{equation}
where $\kappa$ is the average attenuation factor along the dielectric \gls{WG} in dB/m, $\lambda_\mathrm{g}=\frac{\lambda}{\eta_\mathrm{eff}}$ is the guided wavelength, and $\eta_\mathrm{eff}$ is the effective refractive index of the dielectric \gls{WG}. Thus, the channel between the \gls{UE} and the \gls{BS} via a \gls{PA} at $\boldsymbol{\psi}_\mathrm{p}(x_\mathrm{p})$ is  
\begin{equation}
    h(x_\mathrm{p})= \sqrt{\beta_\mathrm{W}(x_\mathrm{p})}h_\mathrm{p}(x_\mathrm{p}). 
\end{equation}
The received signal at the BS is $y=h(x_\mathrm{p})s+n$, where $s$ is the transmitted signal with $\mathbb{E}[|s|^2]=E_s$, and $n\sim\mathcal{CN}(0,\sigma^2)$ is the additive white Gaussian noise. Therefore, the optimal instantaneous \gls{SNR} is
\begin{equation}
    \mathrm{SNR^{opt}}=\sup_{0\leq x_\mathrm{p}\leq L}\{\beta(x_\mathrm{p})|\tilde{h}_\mathrm{p}(x_\mathrm{p})|^2\},
\end{equation}
where $\beta(x_\mathrm{p})= \frac{E_s}{\sigma^2}|\beta_\mathrm
W(x_\mathrm{p})|\beta_\mathrm{LS}(x_\mathrm{p})$. In contrast, the long-term optimal \gls{SNR} is dependent only on \gls{LSF} and \gls{WG} losses,
\begin{equation}
    \mathrm{SNR_{LT}^{opt}}= \sup_{0\leq x_\mathrm{p}\leq L}\left\{\beta(x_\mathrm{p})\right\}.
\end{equation}
The position at which $\mathrm{SNR_{LT}^{opt}}$ occurs is referred to as $x_{\mathrm{p,LT}}^{\mathrm{opt}}$. Although determining the globally optimal \gls{PA} position is desirable, doing so over a continuous \gls{WG} with a highly oscillatory  instantaneous \gls{SNR} response is computationally prohibitive. Instead, we adopt the accurate \gls{TS} approximation  \cite{inwood_analysis_2026}. First, the long-term optimal position, $x_{\mathrm{p,LT}}^{\mathrm{opt}}$, and the corresponding \gls{SNR}, $\mathrm{SNR_{LT}^{opt}}$, are determined. Second, the best of $N$ independent \gls{SSF} samples in the vicinity of $x_{\mathrm{p,LT}}^{\mathrm{opt}}$ is selected.
As $\beta(x_\mathrm{p})$ and $\tilde{h}_\mathrm{p}(x_\mathrm{p})$ are independent, the \gls{SNR} obtained using the \gls{TS} approach is
\begin{equation}
    \mathrm{SNR}^\mathrm{opt}\approx \mathrm{SNR}_\mathrm{TS}=\mathrm{SNR_{LT}^{opt}}|\tilde{h}_\mathrm{p}^\mathrm{max}|^2, \label{eq:SNRopt}
\end{equation}
where $|\tilde{h}_\mathrm{p}^\mathrm{max}|^2=\max_{1\leq n\leq N}\{|\tilde{h}_\mathrm{p}(x_{\mathrm{p},n})|^2\}$ is the maximum fading channel gain among the $N$ equally-spaced points sampled between $x_\mathrm{p,LT}^\mathrm{opt}-\Delta$ and $x_\mathrm{p,LT}^\mathrm{opt}+\Delta$, $\Delta=\delta(N-1)/2$, and $\delta\geq d_\mathrm{SS}$ is the distance between independent samples. This approach enables the effects of large- and small-scale fading to be isolated and analyzed separately, providing greater insight into their respective contributions while offering an efficient and straightforward method for position selection.
\vspace{0.5em}

\section{Analysis Under PA Position Errors}
\label{sec:PAposerr}
\vspace{0.25em}

Using the \gls{TS} approach detailed in Sec.~\ref{sec:sys_mod}, the first stage identifies 
$x_\mathrm{p,LT}^\mathrm{opt}$, the position offering the optimal long-term \gls{SNR}. In practice, however, the \gls{PA} cannot always be placed exactly at this point. This can be for a number of reasons, such as discrete \gls{PA} positions or mechanical tolerances. This section quantifies the resulting performance loss. We model the long-term SNR loss process as a function of a Brownian meander, and use this to derive accurate approximations to its density, distribution, and mean.

\subsection{Density of the Loss}

The exponentially correlated Gaussian process $X(x_\mathrm{p})$ driving the shadow fading can be conveniently represented in terms of a standard \gls{BM}, as shown in \cite{inwood_analysis_2026}, where
\begin{equation}
    X(x_\mathrm{p})=\mathrm{e}^{-\alpha x_\mathrm{p}}B\left(\mathrm{e}^{2\alpha x_\mathrm{p}}\right), \label{eq:BM_transform}
\end{equation}
and $B(t)\sim\mathcal{N}(0,t)$ denotes standard \gls{BM}. Applying this gives
\begin{equation}
    \log(\beta(x_\mathrm{p})) =  C(x_\mathrm{p}) + \sigma_X\mathrm{e}^{-\alpha x_\mathrm{p}}B(\mathrm{e}^{2\alpha x_\mathrm{p}}),
\end{equation}
where $C(x_\mathrm{p})\!=\!\log((E_s/\sigma^2)A) \!+\!\log(|\beta_\mathrm{W}(x_\mathrm{p})|)-\gamma\log(d(x_\mathrm{p}))$ and $\sigma_X\!=\!\sigma_\mathrm
{SF}\log(10)/10$. Changing to Brownian time such that $t_\mathrm{b}=\mathrm{e}^{2\alpha x_\mathrm{p}}$, and thus $x_\mathrm{p}=\log(t_\mathrm{b})/(2\alpha)=\Omega(t_\mathrm{b})$,
\begin{equation}
    \log(\beta(\Omega(t_\mathrm{b})))=C(\Omega(t_\mathrm{b})) + \tfrac{\sigma_X}{\sqrt{t_\mathrm{b}}}B(t_\mathrm{b}).
\end{equation}
The loss due to positioning error can be interpreted as the difference between the process maximum and its value at a given distance $d_\mathrm{err}=|x_\mathrm{p,LT}^\mathrm{opt}-x_\mathrm{p}|$ from the maximizing point, i.e., $\Lambda(t_\mathrm{b})=\log(\beta(\Omega(t^\star)))-\log(\beta(\Omega(t_\mathrm{b})))$, where $t^\star=\mathrm{e}^{2\alpha x_\mathrm{p,LT}^\mathrm{opt}}$. The behavior of standard \gls{BM} around its maximum is exactly characterized by a Brownian meander \cite{denisov_random_1984}. The loss is given by \gls{BM} conditioned on remaining positive, which is appropriate since the process cannot exceed its global maximum. For $\tau_0=\sup\{t\in[0,1]; B(t)=0\}$, a Brownian meander on the interval $[0,1]$ can be defined as \cite{denisov_random_1984}
\begin{equation}
    B^+(t)=\tfrac{1}{\sqrt{1-\tau_0}}|B(\tau_0+t(1-\tau_0))|, \quad 0\leq t \leq 1.
\end{equation}
We now use the Brownian meander as a method to approximate the loss due to positioning error. In this case, once we have conditioned on the fact that the peak occurred at $t^\star$, we only care about the behavior after that point\footnote{Without loss of generality, we assume the selected position occurs after the peak. If the selected position occurs before the peak, the process can be reversed, after which the same method applies.}. Thus, let $u=t_\mathrm{b}-t^\star$ denote the Brownian time past the peak, so that $u=0$ corresponds to $t_\mathrm{b}=t^\star$. Since $C(\Omega(t^\star +u))$ is non-linear in $u$ and positioning errors are likely to be small, it is reasonable to approximate $C(\Omega(t^\star+u))$ by its first-order Taylor expansion about the peak and to treat the variance scaling $\sigma_X/\sqrt{t_\mathrm{b}}$ as approximately constant at $\sigma_X/\sqrt{t^\star}$ over the values of $u$ considered. Therefore, near the peak, we model $\Lambda(u)$ as a Brownian meander with linear drift:
\begin{equation}
    \Lambda(u) \approx \sigma_0(\mu u + \mathcal{M}(u)), \qquad 0\leq u \leq T,  \label{eq:meander_drift}
\end{equation}
where $\mathcal{M}$ is a standard Brownian meander on $[0,T]$, obtained from the unit meander via $\mathcal{M}(u) = \sqrt{T}\,B^+(u/T)$,  $T\leq \mathrm{e^{2\alpha L}-t^\star}$, $\mu=\mu_0/\sigma_0$, $\sigma_0 = \sigma_X/\sqrt{t^\star}$ and
\begin{equation}
    \mu_0 = \tfrac{\mathrm{d}}{\mathrm{d}u}C(\Omega(t^\star+u))\big|_{u=0}.
\end{equation}
The expression for $\mu_0$ can be solved via the chain rule, giving
\begin{equation}
    \mu_0= -\tfrac{1}{2\alpha t^\star}\Big(\tfrac{\kappa\log(10)}{20}+\gamma\tfrac{x_\mathrm{p,LT}^\mathrm{opt}-x_\mathrm{u}}{d(x_\mathrm{p,LT}^\mathrm{opt})^2}\Big).
\end{equation}
The density of a Brownian meander with drift is defined in \cite[Eq. (1.4)]{iafrate_some_2020}. Applying this gives the density of $\Lambda(u)$ as
\begin{equation}
    f_\Lambda(y) = \tfrac{1}{\sigma_0}\tilde{f}_\Lambda\big(\tfrac{y}{\sigma_0}\big),
\end{equation}
where
\begin{equation}
    \tilde{f}_\Lambda\left(z\right)=\left(\tfrac{T}{s}\right)^{\!3/2}\tfrac{ze^{-z^2/(2s)}}{\sqrt{2\pi(T-s)}}\tfrac{G(z)}{D(T)}, \label{eq:ftildeLambday}
\end{equation}
and $s=\mathrm{e}^{2\alpha x_\mathrm{p,LT}^\mathrm{opt}}(\mathrm{e}^{2\alpha d_\mathrm{err}}-1)$. Note that all $y$ terms and the expression as a whole have been scaled by $1/\sigma_0$, as the original expression in \cite{iafrate_some_2020} was derived assuming unit variance. Letting $\upsilon=T-s$, we can write \cite{iafrate_some_2020}
\begin{equation}
    G(z) \!= \!\!\int_0^\infty\!\!\Big(\!\mathrm{e}^{-\frac{(w-z)^2}{2\upsilon}+\mu w}\!-\!\mathrm{e}^{-\frac{(w+z)^2}{2\upsilon}+\mu w}\!\Big)dw,
\end{equation}
\begin{equation}
    D(T) = \int_0^\infty w\mathrm{e}^{-w^2/(2T)+\mu w}dw.
\end{equation}
The integrals in $G(z)$ and $D(T)$ are solved in Appendix A,
\begin{equation}
    G(z)= \!\sqrt{2\pi\upsilon}\,\mathrm{e}^{\frac{\upsilon\mu^2}{2}}\!\Big(\!\mathrm{e}^{\mu z}\Phi\!\left(\!\tfrac{z+\mu\upsilon}{\sqrt{\upsilon}}\!\right)\!- \mathrm{e}^{-\mu z}\Phi\!\left(\!\tfrac{\mu\upsilon-z}{\sqrt{\upsilon}}\!\right)\!\!\!\Big)\!, \label{eq:Gy}
\end{equation}
\begin{equation}
    D(T) = T\big(1+\theta_T\tfrac{\Phi(\theta_T)}{\varphi(\theta_T)}\big), \label{eq:Dt}
\end{equation}
where $\theta_T=\mu\sqrt{T}$. Thus, the complete \gls{PDF} is 
\begin{multline}
    f_\Lambda(y)=\tfrac{\sqrt{T}}{\sigma_0^2s\sqrt{s}}\tfrac{y\,\mathrm{e}^{\upsilon\mu^2/2-y^2/(2s\sigma_0^2)}}{(1+\theta_T\Phi(\theta_T)/\varphi(\theta_T))} \\\times\Big(\mathrm{e}^{\frac{\mu y}{\sigma_0}}\Phi\!\left(\tfrac{\frac{y}{\sigma_0}+\mu\upsilon}{\sqrt{\upsilon}}\right)  - \mathrm{e}^{-\frac{\mu y}{\sigma_0}}\Phi\!\left(\tfrac{\mu\upsilon-\frac{y}{\sigma_0}}{\sqrt{\upsilon}}\right)\!\!\Big). \label{eq:fLambday}
\end{multline}
\begin{rem}
    The parameter $\mu$ characterizes how rapidly the deterministic channel gain changes around the optimal \gls{PA} position, relative to the shadow fading variability. When this varies sharply with position, even a small error can have a noticeable effect on the received signal. As $\mu$ appears throughout  \eqref{eq:fLambday}, it influences the shape and mean of \eqref{eq:fLambday}.
\end{rem}
\subsection{Distribution of the Loss}
As the density is only defined for $y>0$,
\begin{align}
    F_\Lambda(y) &= P(\Lambda(u)\leq y) = \int_0^y f_\Lambda\left(z\right)dz, \quad z\geq 0, \notag 
\end{align}
\begin{align}
    F_\Lambda(y)&= \!\tfrac{\sqrt{T}}{s\sqrt{s}}\tfrac{\mathrm{e}^{\upsilon \mu^2/2}}{\sigma_0(1\!+\!\theta_T\Phi(\theta_T)/\varphi(\theta_T))}\!\int_0^y\!\!\tfrac{z}{\sigma_0}\mathrm{e}^{\frac{-z^2}{2s\sigma_0^2}}\!\Big(\!\mathrm{e}^{\frac{\mu z}{\sigma_0}} \notag  \\ &\qquad\qquad\quad \times\Phi\!\Big(\!\tfrac{\frac{z}{\sigma_0}+\mu\upsilon}{\sqrt{\upsilon}}\!\Big)\! -\! \mathrm{e}^{\frac{-\mu z}{\sigma_0}}\Phi\!\Big(\!\tfrac{\mu\upsilon-\frac{z}{\sigma_0}}{\sqrt{\upsilon}}\!\Big)\!\!\Big)dz.
\end{align}
The integral over $z$ is solved in Appendix B, leading to the final result in \eqref{eq:FLambday} at the top of the next page with $\theta_x\!=\!\mu\sqrt{x}$, $\Xi_x^+(y)\!=\! \frac{y}{\sigma_0\sqrt{x}}\!+\!\theta_x$, $\Xi_x^-(y)\!=\!\frac{y}{\sigma_0\sqrt{x}}\!-\!\theta_x$ and $\rho_s\!=\!-\sqrt{s/T}$.

\begin{figure*}[!t]
\begin{multline}
     F_\Lambda(y)\! =\!\tfrac{\sqrt{T}\mathrm{e}^{\upsilon\mu^2/2}}{\sqrt{s}(1\!+\!\theta_T\Phi(\theta_T)/\varphi(\theta_T))}\!\Big[\mathrm{e}^{\!-\frac{y^2}{2s\sigma_0^2}-\frac{y\mu}{\sigma_0}}\!-\!1\!+\mathrm{e}^{\frac{\theta_s^2}{2}}\!\sqrt{2\pi}\Big[\theta_s\!\left(\Phi(\Xi^+_s\!(y)\!)\!-\!\Phi\!\left(\theta_s\right)\!\right)\!-\!\Phi(\Xi^+_\upsilon\!(y)\!)\varphi(\Xi^-_s\!(y)\!)\! -\!\Phi(\Xi^-_\upsilon\!(y)\!)\varphi(\Xi_s^+\!(y)\!)\!+\!\sqrt{\!\tfrac{s}{s+\upsilon}}\\\times\! \varphi(\theta_{T})\!\Big(\!2\Phi\Big(\!\tfrac{y}{\sigma_0}\!\sqrt{\!\tfrac{s+\upsilon}{s\upsilon}}\Big)\!-\!1\!\Big)\!+\!\varphi(\theta_s)\!+\!\theta_s\!\big(\!\Phi_2(\Xi_s^-\!(y);\!\theta_{T},\! \rho_s\!)\!-\!\Phi_2(\!-\theta_s;\!\theta_{T},\!\rho_s\!)\!-\!\Phi_2(\Xi_s^+\!(y);\!-\theta_{T},\! \rho_s\!)\!+\!\Phi_2(\theta_s;\!-\theta_{T},\!\rho_s\!)\!\big)\!\Big]\!\Big]\!. \label{eq:FLambday}
\end{multline}
\hrulefill
\end{figure*}

\begin{rem}
    The CDF in \eqref{eq:FLambday} depends on $\rho_s=-\sqrt{s/T}$, linking the positioning error to the remaining \gls{WG} length after the optimal \gls{PA} location. This indicates that a position error may produce different loss distributions depending on how close the optimal \gls{PA} position is to the end of the waveguide.
\end{rem}

\subsection{Mean Loss}
Applying \eqref{eq:ftildeLambday} multiplied by $\sigma_0$ (as \eqref{eq:ftildeLambday} is normalized),
\begin{align}
    \mathbb{E}[\Lambda(u)]&=\sigma_0\int_0^\infty y\tilde{f}_\Lambda(y)dy, \notag \\
    & = \tfrac{\sqrt{T}}{s\sqrt{s}}\tfrac{\sigma_0\mathrm{e}^{\upsilon \mu^2/2}}{(1+\theta_T\Phi(\theta_T)/\varphi(\theta_T))}\!\int_0^\infty\!y^2\mathrm{e}^{-\frac{y^2}{2s}} \notag  \\ &\qquad\qquad\times\Big(\!\mathrm{e}^{\mu y}\Phi\!\left(\!\tfrac{y+\mu\upsilon}{\sqrt{\upsilon}}\!\right)\! -\! \mathrm{e}^{-\mu y}\Phi\!\left(\!\tfrac{\mu\upsilon-y}{\sqrt{\upsilon}}\!\right)\!\!\!\Big)dy. \label{eq:ELambdau_ints}
\end{align}
This expression is solved in Appendix B, leading to
\begin{align}
    \mathbb{E}[\Lambda(u)]\!&=\tfrac{\sqrt{T}}{s\sqrt{s}}\tfrac{\sigma_0\mathrm{e}^{\upsilon \mu^2/2}}{(1+\theta_T\Phi(\theta_T)/\varphi(\theta_T))}\Big[\mathrm{e}^{\frac{\mu^2s}{2}}s\sqrt{2\pi s}\Big(\!\!(1\!+\!\theta_s^2) \notag \\ &\times\!(\Phi_2(\theta_{T};\theta_s,\!-\rho_s)\!+\!\Phi_2(\!-\theta_{T};\!-\theta_s,\!-\rho_s)\!-\!1\!+\!\Phi(\theta_s)\!) \notag \\ & +\!\theta_s\varphi(\theta_s)(2\Phi(\theta_\upsilon)\!-\!1)\!-\!\rho_s\sqrt{\!\tfrac{2\upsilon}{\pi}}\varphi(\theta_{T})\!\!\Big)\!\!+\!s^2\mu\Big]\!.\! \label{eq:ELambdau}
\end{align}
The expressions in \eqref{eq:fLambday}, \eqref{eq:FLambday} and \eqref{eq:ELambdau} are later employed in Sec. \ref{sec:numres} to study \gls{PASS} performance.
\vspace{0.25em}
\section{Analysis Under Imperfect CSI}
\vspace{0.25em}
\label{sec:imperf_CSI}
A common imperfection in many systems is channel estimation error, which can arise from many sources including additive noise, pilot contamination and synchronization errors. Therefore, the impacts of this error are important to quantify for a \gls{PASS} under fading channels. In this section, we derive the mean and distribution of the \gls{SCP} for a \gls{PA} at the best position selected from noisy channel estimates for a system employing the \gls{TS} approach. As we assume that $\beta_\mathrm{LS}(x_\mathrm{p})$ and $h_\mathrm{LoS}(x_\mathrm{p})$ vary slowly and can be estimated statistically, we focus on the losses resulting from imperfect estimation of $h_\mathrm{SC}(x_\mathrm{p})$. This provides a robust analysis of the worst-case scenario, as the stability of the deterministic \gls{LoS} component reduces the effects of imperfect \gls{CSI}.

\subsection{Mean and Variance of the SCP Under Imperfect CSI}
Consider $N$ instances of $h_\mathrm{SC}(x_\mathrm
p)$ in the vicinity of $x_\mathrm{p,LT}^\mathrm{opt}$ as discussed in Sec. \ref{sec:sys_mod}, where $x_\mathrm{p}\in\{x_\mathrm{p,LT}^\mathrm{opt}-\Delta+k\delta,\,\, k=0,\dots ,N-1\}$ with $\Delta=\delta(N-1)/2$, and $\delta\geq d_\mathrm{SS}$ is the distance between independent samples. For simplicity, let us denote these channels in this section as $h_1,\dots, h_N$, with powers $Y_1 = |h_1|^2,\dots, Y_N = |h_N|^2$. Without loss of generality, the channels are assumed to be normalized, such that $Y_i\sim \mathrm{Exp}(1)$. The received signal is measured $T_\mathrm{s}$ times at each position, so at position $i$,
\begin{equation}
    [x_{i1}, \dots, x_{iT_\mathrm{s}}]^T
    = h_i s + [n_{i1}, \dots, n_{iT_\mathrm{s}}]^T,
\end{equation}
where $n_i\sim\mathcal{CN}(0,\sigma^2)$ is \gls{AWGN} and $s$ is the transmitted signal with $|s|=1$. This leads to an estimated channel of
\begin{equation}
    \hat{h}_i=\tfrac{1}{sT_\mathrm{s}}\sum_{l=1}^{T_\mathrm{s}}x_{il}=h_i+\tfrac{1}{sT_\mathrm{s}}\sum_{l=1}^{T_\mathrm{s}}n_{il}=h_i + z_i,
\end{equation}
where $z_i\!\sim\! \mathcal{CN}(0,\sigma^2/T_\mathrm{s})$. Hence, $\hat{h}_i\!\sim\!\mathcal{CN}(0,1\!+\!\sigma^2/T_\mathrm{s})$ and the corresponding estimated power is $X_i\!=\!|\hat{h}_i|^2\!\sim\!\mathrm{Exp}((1\!+\!\sigma^2/T_\mathrm{s})^{-1})$. Let $\lambda_T\!=\!(1\!+\!\sigma^2/T_\mathrm{s})^{-1}$, leading to $X_i\!\sim\!\mathrm{Exp}(\lambda_T).$ Thus, the $X_i$ terms denote the measured channel, and $Y_i$ terms denote the true channel. Let $X_\mathrm{m}\!=\!\max(X_i)$ denote the best observed \gls{SSF} instance.  The corresponding true power at that location, $Y_\mathrm{m}$, is a concomitant \cite{david_order_2003}. This may not be the true optimal due to estimation error, so the mean loss is
\begin{equation}
    \mathbb{E}[Y_{\mathrm{loss}}]=\mathbb{E}[Y_{\mathrm{max}}] - \mathbb{E}[Y_{\mathrm{m}}], \label{eq:EYloss}
\end{equation}
where, from \cite[Eq. (6.5)]{feller_introduction_1971},
\begin{equation}
    \mathbb{E}[Y_\mathrm{max}]= \mathbb{E}[\max(Y_i)]=H_N, \label{eq:EYmax}
\end{equation}
with $H_N=\sum^N_{k=1}1/k$ denoting the standard harmonic number. The required first moment of $Y_\mathrm{m}$ is given in the following lemma, alongside its second moment and variance.
\begin{lemma}
    The mean of $Y_\mathrm{m}$ is
    \begin{equation}
    \mathbb{E}[Y_\mathrm{m}] = 1 + \lambda_T(H_N-1), \label{eq:EYm}
\end{equation}
and the second moment of $Y_\mathrm{m}$ is
\begin{equation}
    \mathbb{E}\!\!\left[Y_\mathrm{m}^2\right]\!=\! 2\!+\! 4\lambda_T(H_N\!-\!1)\!+\!\lambda_T^2\big(\!H_N^2\!-\!4H_N\!+\!2\!+\!H_N^{(2)}\big)\!, \label{eq:EYm2}
\end{equation}
where $H_N^{(2)}\!\!=\!\sum^N_{k=1}\!\frac{1}{k^2}$ is the generalized harmonic number of order 2. The variance of $Y_\mathrm{m}$, $\mathrm{Var}[Y_\mathrm{m}]\!=\!\mathbb{E}\!\left[Y_\mathrm{m}^2\right]\!-\!\mathbb{E}[Y_\mathrm{m}]^2$, is
\begin{equation}
    \mathrm{Var}[Y_\mathrm{m}]= 1\!+\!2\lambda_T(H_N\!-1)\!+\!\lambda_T^2\big(H_N^{(2)}\!-\!2H_N\!+\!1\big).
\end{equation}
\end{lemma}
\begin{proof}
    See Appendix C.
\end{proof}
\begin{rem}
    The expression in \eqref{eq:EYm} clearly shows the relationship between $Y_\mathrm{m}$ and $X_\mathrm{m}$, and thus the effects of channel estimation quality. As $\mathbb{E}[X_\mathrm{m}]=H_N/\lambda_T$, we can write
    \begin{equation}
        \mathbb{E}[Y_\mathrm{m}]=1-\lambda_T+\lambda_T^2\mathbb{E}[X_\mathrm{m}].
    \end{equation}
    As $\lambda_T\to 1$, which can occur when the noise power is very low ($\sigma^2\to 0)$ or a large number of samples are taken ($T_\mathrm{s}\to \infty$), $\mathbb{E}[Y_\mathrm{m}]\to\mathbb{E}[X_\mathrm{m}]$. This indicates perfect \gls{CSI}, and accordingly, no loss due to channel estimation error. In contrast, when $\lambda_T\to 0$, which can occur as $\sigma^2\to \infty$ or $T_\mathrm{s}\to 0$, $X_i$ and $Y_i$ become effectively uncorrelated, and thus selection is essentially random. Here, increasing $N$ will provide no improvement.
\end{rem}
\subsection{Distribution of the SCP Under Imperfect CSI}
The distribution of the true \gls{SCP} at the location corresponding to the maximum observed \gls{SCP} is expressed below.
\begin{lemma}
    The distribution of $Y_\mathrm{m}$ is 
    \begin{equation}
        P(Y_\mathrm{m}<y)=1\!-\!N\sum_{j=0}^{N-1}\!\tbinom{N-1}{j}\tfrac{(-1)^j}{j+1}\exp\!\left(\!-\tfrac{(j+1)y}{1+j(1-\lambda_T)}\!\right)\!\!. \label{eq:PYmlty}
    \end{equation}
\end{lemma}
\begin{proof}
    See Appendix C.
\end{proof}
\begin{rem}
The CDF of $Y_\mathrm{m}$ reveals a clean structural relationship to the perfect-CSI benchmark. Defining $\mu_j\!=\!1\!+\!j(1\!-\!\lambda_T)$,
\begin{equation}
    P(Y_\mathrm{m}<y) = 1-N\sum_{j=0}^{N-1}\!\tbinom{N-1}{j}\tfrac{(-1)^j}{j+1}\mathrm{e}^{-\frac{j+1}{\mu_j}y}\!, \label{eq:PYmly}
\end{equation}
which is identical in form to \cite[Eq. (2.1.1)]{david_order_2003}
\begin{equation}
    P(Y_\mathrm{max}\!\!<\!y)\!\!=\!\! (\!1\!-\mathrm{e}^{-y})^N\!\!=\!1\!-\!N\!\!\sum_{j=0}^{N-1}\!\!\tbinom{N-1}{j}\!\tfrac{(-1)^j}{j+1}\mathrm{e}^{-(j+1)y}. \label{eq:PYmax}
\end{equation}
\end{rem}
\begin{rem}
Notably, the $j=0$ term dominates the sum in \eqref{eq:PYmly} as $y\to\infty$, giving the tail behavior
\begin{equation}
    P(Y_\mathrm{m}<y) \sim 1-N\,\mathrm{e}^{-y}, \qquad y\to\infty,
\end{equation}
identical to that of the perfect-CSI benchmark. Imperfect channel estimation therefore has negligible effect on the probability of an exceptionally favorable position.
\end{rem}
\section{Analysis Under LoS Blockages}
\vspace{0.25em}
\label{sec:blockage}

As \gls{PASS} \glspl{WG} are anticipated to have lengths in the order of meters, it is reasonable to assume that some \gls{PA} positions will not result in an \gls{LoS} channel between the \gls{PA} and \gls{UE}. Therefore, in this section, we analyze a \gls{PASS} where part of the \gls{WG} is obstructed by one or more blockages. We obtain an expression for the distribution of the optimum long-term \gls{SNR} under an arbitrary number of blockages, and evaluate this expression for the special case of a single blockage. This is enabled by expressing the problem in terms of standard \gls{BM}.

\subsection{General Model}
In the case of a \gls{PASS} with no blockages along the \gls{WG}, the long-term \gls{SNR} involves correlated shadow fading driven by a single correlated Gaussian process. The introduction of blocked regions, and thus \gls{NLoS} links, leads to the introduction of a second process with a higher \gls{PLE}. The correlated shadow fading is thus assumed to be driven by $X_\mathrm{L}(x_\mathrm{p})\sim\mathcal{N}(0,1)$ when the link is \gls{LoS} and $X_\mathrm{N}(x_\mathrm{p})\sim\mathcal{N}(0,1)$ when the link is \gls{NLoS}. Both are defined over the full interval $0\leq x_\mathrm{p}\leq L$. We assume that $X_\mathrm{L}(x_\mathrm{p})$ and $ X_\mathrm{N}(x_\mathrm{p})$ are independent of each other but internally correlated over the entire interval.

\begin{figure}[ht]
    \centering
    \includegraphics[width=0.95\linewidth]{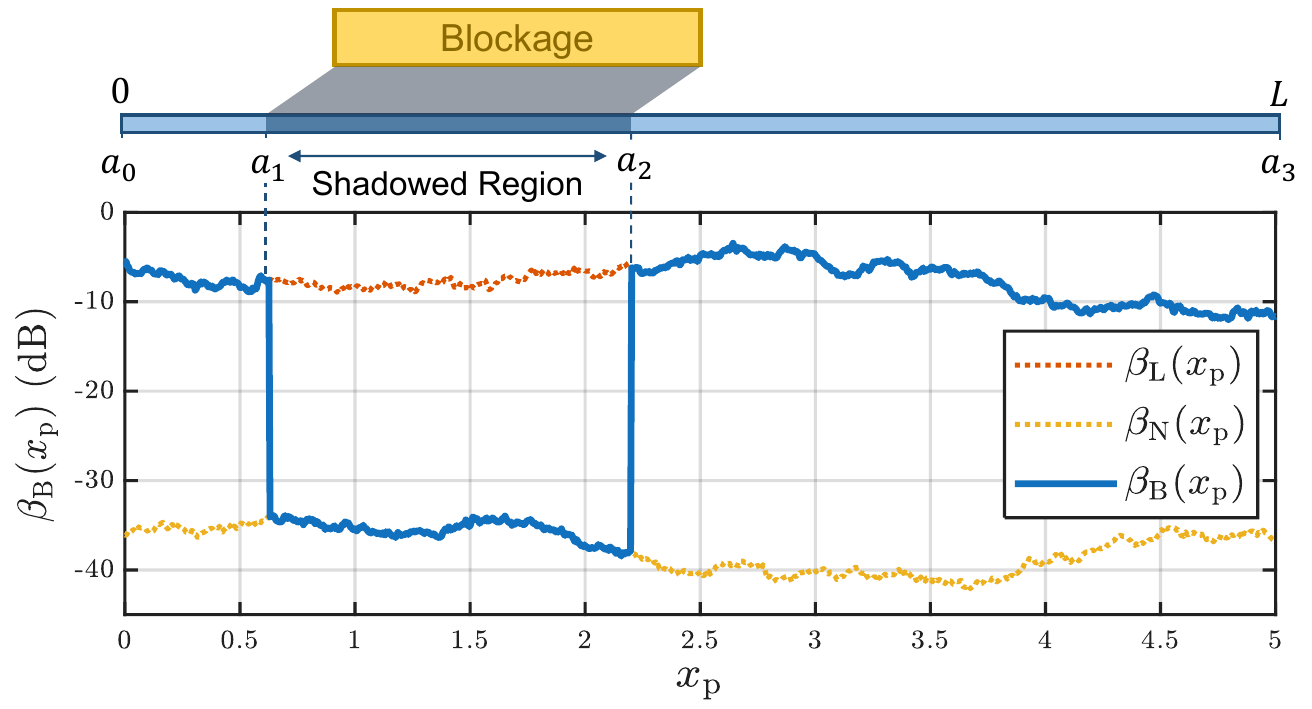}
    \caption{A simulated example of the long-term \gls{SNR} across the \gls{WG} for the case of a single blockage.}
    \label{fig:blockage_model}
\end{figure}

Let there be $M$ blockages, with the switching points $0=a_0\leq a_1<a_2 <\dots <a_{2M} \leq a_{2M+1}=L$. The set of \gls{LoS} regions can thus be defined as $\mathcal{I}_\mathrm{L}=[a_0,a_1]\cup[a_2,a_3]\cup\dots\cup[a_{2M},a_{2M+1}]$ and the set of \gls{NLoS} regions can be defined as $\mathcal{I}_\mathrm{N}=[a_1,a_2]\cup[a_3,a_4]\cup\dots\cup[a_{2M-1},a_{2M}]$. Note that the channel is \gls{LoS} at the start of the \gls{WG} when $a_0<a_1$, and \gls{NLoS} when $a_0=a_1$. The same behavior occurs at the end of the \gls{WG}, with $a_{2M}<a_{2M+1}$ denoting an \gls{LoS} channel and $a_{2M}=a_{2M+1}$ denoting an \gls{NLoS} channel. For a fixed blockage configuration $\mathcal{B}=(a_1,\dots,a_{2M})$, the observed process, $X_\mathrm{obs}(x_\mathrm{p})$, is $X_\mathrm{L}(x_\mathrm{p})$ when $x_\mathrm{p}\in\mathcal{I}_\mathrm{L}$ and $X_\mathrm{N}(x_\mathrm{p})$ when $x_\mathrm{p}\in\mathcal{I}_\mathrm{N}$. This leads to a long-term \gls{SNR} under blockages, $\beta_\mathrm{B}(x_\mathrm{p})$, of $\beta_\mathrm{L}(x_\mathrm{p})$ when $x_\mathrm{p}\in\mathcal{I}_\mathrm{L}$ and $\beta_\mathrm{N}(x_\mathrm{p})$ when $x_\mathrm{p}\in\mathcal{I}_\mathrm{N}$.
Here, $\beta_\mathrm{L}(x_\mathrm{p})=\frac{E_s}{\sigma^2}|\beta_\mathrm
W(x_\mathrm{p})| A\,d(x_\mathrm{p})^{-\gamma_\mathrm{L}}L_\mathrm{L}(x_\mathrm{p})$,  $\beta_\mathrm{N}(x_\mathrm{p})=\frac{E_s}{\sigma^2}|\beta_\mathrm
W(x_\mathrm{p})| A\,d(x_\mathrm{p})^{-\gamma_\mathrm{N}}L_\mathrm{N}(x_\mathrm{p})$, $L_\mathrm{L}(x_\mathrm{p})=10^{\sigma_\mathrm{SF}X_\mathrm{L}(x_\mathrm{p})/10}$, $L_\mathrm{N}(x_\mathrm{p})=10^{\sigma_\mathrm{SF}X_\mathrm{N}(x_\mathrm{p})/10}$ and $\gamma_\mathrm{L}$ and $\gamma_\mathrm{N}$ are the \glspl{PLE} corresponding to the \gls{LoS} and \gls{NLoS} links, respectively. An example $\beta_\mathrm{B}(x_\mathrm{p})$ for a system involving a \gls{UE} at $\boldsymbol{\psi}_\mathrm{u}=(2.5,5)$ communicating with a \gls{PA} on an $L=5$ m \gls{WG} with a single blockage is shown in Fig. \ref{fig:blockage_model}. In this case, $\gamma_\mathrm{L}=2.2$, $\gamma_\mathrm{N}=4.5$, $\sigma_\mathrm{SF}=8$ dB and $f_\mathrm{c}=28$ GHz.
\vspace{-0.8em}
\vspace{-0.5em}

Let $\beta_\mathrm{B}^\star=\sup_{x_\mathrm{p}\in[0,L]}\{\beta_\mathrm{B}(x_\mathrm{p})\}$. The distribution of $\beta_\mathrm{B}^\star$ for blockage configuration $\mathcal{B}$, $F_{\beta_\mathrm{B}^\star}(x\mid\mathcal{B})=P(\beta_\mathrm{B}^\star\leq x\mid\mathcal{B})$, can be obtained by rewriting the expression in terms of $X_\mathrm{obs}(x_\mathrm{p})$,
\begin{equation}
    F_{\beta_\mathrm{B}^\star}(x\mid \mathcal{B})= P(X_\mathrm{obs}(x_\mathrm{p})\leq y_\mathrm{obs}(x_\mathrm{p},x)\cap\forall x_\mathrm{p}\in[0,L]\mid \mathcal{B}), \notag
\end{equation}
where $y_\mathrm{obs}(x_\mathrm{p},x)$ is $y_\mathrm{L}(x_\mathrm{p},x)$ if $x_\mathrm{p}\in\mathcal{I}_\mathrm{L}$ and $y_\mathrm{N}(x_\mathrm{p},x)$ if $x_\mathrm{p}\in\mathcal{I}_\mathrm{N}$. Here, $y_\mathrm{L}(x_\mathrm{p},x)=\frac{10}{\sigma_\mathrm{SF}\log(10)}(\log(x)+\Psi_\mathrm{L}(x_\mathrm{p}))$, $y_\mathrm{N}(x_\mathrm{p},x)=\frac{10}{\sigma_\mathrm{SF}\log(10)}(\log(x)+\Psi_\mathrm{N}(x_\mathrm{p}))$, $\Psi_\mathrm{L}(x_\mathrm{p}){=}\log\!\left(\!\frac{\sigma^2d(x_\mathrm{p})^{\gamma_\mathrm{L}}}{E_sA|\beta_\mathrm{W}(x_\mathrm{p})|}\right)$ and $\Psi_\mathrm{N}(x_\mathrm{p}){=}\log\!\left(\!\frac{\sigma^2d(x_\mathrm{p})^{\gamma_\mathrm{N}}}{E_sA|\beta_\mathrm{W}(x_\mathrm{p})|}\right)$. As $X_\mathrm{L}(x_\mathrm{p})$ and $X_\mathrm{N}(x_\mathrm{p})$ are independent,
\begin{equation}
    F_{\beta_\mathrm{B}^\star}(x\mid \mathcal{B})=F_\mathrm{L}(x\mid\mathcal{B})F_\mathrm{N}(x\mid\mathcal{B}), \label{eq:Fs}
\end{equation}
where 
\begin{align}
    F_\mathrm{L}(x\mid\mathcal{B})&= P(X_\mathrm{L}(x_\mathrm{p})\leq y_\mathrm{L}(x_\mathrm{p},x)\cap\,\forall\, x_\mathrm{p}\in\mathcal{I}_\mathrm{L}),\\
    F_\mathrm{N}(x\mid\mathcal{B})&= P(X_\mathrm{N}(x_\mathrm{p})\leq y_\mathrm{N}(x_\mathrm{p},x)\cap\,\forall\, x_\mathrm{p}\in\mathcal{I}_\mathrm{N}).
\end{align}
As we are considering the distribution of the maximum $\mathrm{SNR_{LT}}$ value, one way to simplify the expression in \eqref{eq:Fs} is to apply the approximation
\begin{equation}
    F_{\beta_\mathrm{B}^\star}(x\mid \mathcal{B})\approx F_\mathrm{L}(x\mid\mathcal{B}).
\end{equation}
For typical \gls{PLE} values, it is very unlikely that the optimal point on the \gls{WG} will occur during a blockage. This concept is illustrated in Fig. \ref{fig:blockage_model}, where all points in the blocked region lead to significantly attenuated \glspl{SNR}. Therefore, we must solve $F_\mathrm{L}(x\mid\mathcal{B})$. As in Sec. \ref{sec:PAposerr}, we can express $X_\mathrm{L}(x_\mathrm{p})$ as a transform of standard \gls{BM}, such that
\begin{equation}
    X_\mathrm{L}(x_\mathrm{p})=\mathrm{e}^{-\alpha x_\mathrm{p}}B_\mathrm{L}(\mathrm{e}^{2\alpha x_\mathrm{p}}).
\end{equation}
Defining $s=\mathrm{e}^{2\alpha x_\mathrm{p}}-1$ and $x_\mathrm{p}=\Omega(s)=\frac{\log(s+1)}{2\alpha}$, the problem now becomes one of $B_\mathrm{L}(s)$ staying below the threshold $g_\mathrm{L}(s,x)=y_\mathrm{L}(\Omega(s),x)\sqrt{s+1}$ in the \gls{LoS} regions $[s_0,s_1],[s_2,s_3],\dots,[s_{2M},s_{2M+1}]$, where $s_i=\mathrm{e}^{2\alpha a_i}-1$. Note that the Brownian time of $B_\mathrm{L}(s)$ is $1+s$ due to the variable transformation. In blocked regions, the \gls{LoS} process is effectively unobserved as $\mathrm{\beta_\mathrm{B}(x_\mathrm{p})}$ follows the \gls{NLoS} process.

The two densities below are used to construct $F_{\beta_\mathrm{B}^\star}(x\mid \mathcal{B})$:
\begin{itemize}
    \item \textbf{Free transition density:} The probability of a process ending at value $w_i$ at $s_i$, given that it started at $w_{i-1}$ at $s_{i-1}$. As $B_\mathrm{L}(s)$ is \gls{BM}, this can be defined as \cite{morters_brownian_2010}
    \begin{equation}
            f_t(w_i\!\mid\! w_{i-1})\!=\!\tfrac{1}{\sqrt{2\pi(s_i-s_{i-1})}}\!\exp\!\left(\!-\tfrac{(w_i-w_{i-1})^2}{2(s_i-s_{i-1})}\!\right)\!\!. \label{eq:ftd}
    \end{equation}
    \item \textbf{First passage density:} The density of when the process first crosses a threshold given it started at $w_i$ at $s_i$. As the threshold is curved in this case, we apply the tangent approximation proposed in \cite{ferebee_tangent_1982}, such that
    \begin{multline}
    f_\tau(\tau\mid w_i) \approx \tfrac{1}{\sqrt{2\pi(\tau-s_i)}}\exp\left(-\tfrac{(g_\mathrm{L}(\tau,x)-w_i)^2}{2(\tau-s_i)}\right) \\ \times \left[\tfrac{g_\mathrm{L}(\tau,x)-w_i}{\tau-s_i}-g'_\mathrm{L}(\tau,x)\right]. \label{eq:TA} 
\end{multline}
\end{itemize}
As a single \gls{LoS} process is employed throughout the \gls{WG}, the joint density of its values across successive regions admits a Markovian factorization. In particular, the conditional density of the process within a given region depends only on its value at the beginning of that region, rather than on the preceding process history. Thus, the joint density of surviving (remaining below the threshold at all points) all regions is
\begin{equation}
    f(w_0,w_1,\dots,w_{2M+1})=f(w_0)\prod_{i=1}^{2M+1}f(w_i\mid w_{i-1}),
\end{equation}
where $f(w_0)=\varphi(w_0)$ and $f(w_i\mid w_{i-1})$ can be constructed from $f_t(\cdot)$ and $f_\tau(\cdot)$, as will now be detailed. Different densities are needed for \gls{LoS} and \gls{NLoS} regions. In \gls{NLoS} regions, we can directly apply the free transition density in \eqref{eq:ftd}, as we are unconcerned as to whether or not $B_\mathrm{L}(s)$ crosses the threshold, only that the process begins at $B_\mathrm{L}(s_{i-1})= w_{i-1}$ and ends at $B_\mathrm{L}(s_i)= w_{i}$.

The free transition density can be thought of as containing all possible paths from $(s_{i-1},w_{i-1})$ to $(s_i,w_i)$ - those that cross the threshold $g_\mathrm{L}(\tau,x)$, and those that do not. In the \gls{LoS} regions, we need the non-crossing transition density. This can be obtained by subtracting the contributions of the paths that cross from the free transition density. Suppose the first crossing occurs at $\tau$. Before $\tau$, the process is below the threshold, and once it has crossed, it no longer needs to be observed as it has not survived the region. Therefore, the contribution of a path that first crosses at $\tau$ can be expressed as $f_\tau(\tau\mid w_{i-1}, s_{i-1})f_t(w_i,s_i\mid g_\mathrm{L}(\tau, x), \tau)$. Integration over $\tau\in[s_{i-1},s_i]$ then generalizes this to all possible first crossing times in the region. Therefore, the non-crossing transition density can be defined as
\begin{multline}
    K_\mathrm{L}(w_i\mid w_{i-1};x) = f_t(w_i\mid w_{i-1}) \\ -\int_{s_{i-1}}^{s_i}f_\tau(\tau\mid w_{i-1})f_t(w_i\mid g_\mathrm{L}(\tau,x))d\tau. \label{eq:KL}
\end{multline}
As all $w_i$ values are random, they must be integrated over. This leads to the general expression for $F_\mathrm{L}(x\mid\mathcal{B})$ of
\begin{align}
    F_\mathrm{L}(x\mid\mathcal{B})=&\int_{-\infty}^{g_\mathrm{L}(s_{2M+1},x)}\!\!\!\int_{-\infty}^{g_\mathrm{L}(s_{2M},x)}\!\!\cdots\!\int_{-\infty}^{g_\mathrm{L}(s_1,x)}\!\!\!\int_{-\infty}^{g_\mathrm{L}(0,x)} \notag \\ &\times \varphi(w_0)\prod_{i=1}^{2M+1}\mathcal{L}_i\, dw_0dw_1\dots dw_{2M}dw_{2M+1}, \notag \\
    \label{eq:genblock}
    =&\int_{\mathcal{D}_{2M+1}}\varphi(w_0)\prod_{i=1}^{2M+1}\mathcal{L}_i\,d\mathbf{w},
\end{align}
where $\mathcal{D}_{2M+1}$ is the region where $-\infty<w_i\leq g(s_i,x)$ for $\mathbf{w}=(w_0,w_1,\dots w_{2M+1})$ and 
\begin{equation}
    \mathcal{L}_i=\begin{cases}
        K_\mathrm{L}(w_i\mid w_{i-1}; x), & i\equiv 1 \pmod{2}, \\
        f_t(w_i \mid w_{i-1}), & i\equiv 0 \pmod{2}.
    \end{cases}
\end{equation}
While $F_\mathrm{L}(x\mid\mathcal{B})$ is typically  sufficient to accurately approximate $F_{\beta_\mathrm{B}^\star}(x\mid\mathcal{B})$ for standard parameter values, for completeness, the same process can be followed to derive $F_\mathrm{N}(x\mid\mathcal{B})$ if desired.

\subsection{Single Blockage Case}
We now focus on the special case involving a single blockage, an example of which is shown in Fig. \ref{fig:blockage_model}. This scenario allows the complex expression in \eqref{eq:genblock} to collapse to one involving only two integrals, which can be easily solved numerically. This simpler form also enables system insights to be obtained. In this case, \eqref{eq:genblock} becomes
\begin{multline}
    F_{\beta_{\mathrm{B}1}^\star}(x\mid \mathcal{B})=\int_{\mathcal{D}_3} \varphi(w_0) K_\mathrm{L}(w_1\mid w_0;x)f_t(w_2\mid w_1) \\ \times K_\mathrm{L}(w_3\mid w_2;x) \,d\mathbf{w}.
\end{multline}
Substituting in the values of $K_\mathrm{L}(\cdot)$ from \eqref{eq:KL} and expanding the  brackets leads to
\begin{equation}
    F_{\beta_{\mathrm{B}1}^\star}(x\mid \mathcal{B})= I_1 -I_2 - I_3 + I_4, \label{eq:FBstar}
\end{equation}
\begin{equation}
    I_1\!=\!\! \int_{\mathcal{D}_3}\varphi(w_0) f_t(w_1\mid w_0)f_t(w_2\mid w_1) f_t(w_3\mid w_2) d\mathbf{w}, \label{eq:I1}
\end{equation}
\begin{multline}
    I_2 = \int_{\mathcal{D}_3}\int^{s_3}_{s_2}\varphi(w_0) f_t(w_1\mid w_0)f_t(w_2\mid w_1)  \\ \times f_\tau(\tau_2\mid w_2) f_t(w_3\mid g_\mathrm{L}(\tau_2, x)) d\tau_2 d\mathbf{w},
\end{multline}
\begin{multline}
    I_3 = \int_{\mathcal{D}_3}\int^{s_1}_{0}\!\varphi(w_0)f_\tau(\tau_1\mid w_0)f_t(w_1\mid g_\mathrm{L}(\tau_1,x))\\ \times f_t(w_2\mid w_1) f_t(w_3\mid w_2) d\tau_1 d\mathbf{w},
\end{multline}
\begin{multline}
    I_4 = \int_{\mathcal{D}_3}\!\int^{s_1}_{0}\!\!\int^{s_3}_{s_2} \varphi(w_0) f_\tau(\tau_1\mid w_0)f_t(w_1\mid g_\mathrm{L}(\tau_1,x))\\ \times\!f_t(w_2\! \mid \! w_1)  f_\tau(\tau_2\! \mid \! w_2)  f_t(w_3\! \mid \! g_\mathrm{L}(\tau_2, x)) d\tau_2 d\tau_1 d\mathbf{w}.
\end{multline}
The expression in \eqref{eq:I1} is a multivariate normal \gls{CDF},
\begin{equation}
    I_1 = \Phi_4\left([b_0,b_1,b_2,b_3]; \mathbf{0}, \mathbf{\Sigma^a}\right), \label{eq:I1_res}
\end{equation}
where $b_i=g_\mathrm{L}(s_i,x)$, $\mathbf{\Sigma}_{ij}^\mathbf{a}$ is a covariance matrix with elements ${\Sigma}_{ij}^{a}=\min(t_i,t_j)$ and $t_i=1+s_i$ (as the process starts from a standard normal \gls{RV}), with $i,j\in[0,3]$.
\begin{rem}
The expression in \eqref{eq:I1_res} gives the probability that the process lies below the threshold at each of the four region boundaries, without accounting for the possibility that the process exceeds the threshold at some point between two consecutive boundaries. This is typically the dominant of the four terms, particularly as the spacing between consecutive boundaries, $s_i-s_{i-1}$, decreases. Intuitively, a shorter interval leaves the process less time to wander above the threshold and return before the next boundary is reached. Consequently, as the boundary spacing shrinks, evaluating the process only at the boundaries rather than continuously becomes an increasingly accurate representation of the entire path, and the correction terms $I_2$, $I_3$ and $I_4$ correspondingly vanish.
\end{rem}
The integrals over $w_i$ in $I_2$, $I_3$ and $I_4$ are solved in Appendix D, with their results stated below.
\begin{equation}
    I_2 = \int_{s_2}^{s_3}D(\tau_2)E(\tau_2)d\tau_2, \label{eq:I2_res}
\end{equation}
where $D(\tau_2) = \Phi\left(\frac{b_3-g_\mathrm{L}(\tau_2,x)}{\sqrt{s_3-\tau_2}}\right)$,
\begin{align}
    &E(\tau_2)\!=\!N(g_\mathrm{L}(\tau_2,x);0,t_f)\Big[\!\left(\tfrac{g_\mathrm{L}(\tau_2,x)-{\mu}_2^{E}}{\tau_2-s_2}-g_\mathrm{L}'(\tau_2,x)\right) \notag \\ &\times\! \Phi_3\left(\![b_0,b_1,b_2];\boldsymbol{\mu}^\mathrm{E}, \boldsymbol{\Sigma}^\mathrm{E}\!\right)\!+\!\tfrac{1}{\tau_2-s_2}\!\sum_{j=0}^2\Sigma^{E}_{2j}N(b_j;\mu^E_j\!,\Sigma^E_{jj}) \notag \\ & \times\! \Phi_2\left(b_{-j}; \boldsymbol{\mu}^\mathrm{E}_{-j\mid j}, \boldsymbol{\Sigma}^\mathrm{E}_{-j\mid j}\right)\!\Big], \label{eq:Etau2}
\end{align}
$t_f=1+\tau_2$, $\boldsymbol{\mu}^\mathrm{E}$ is a mean vector with elements $\mu_i^E=\frac{t_i}{t_f}g_\mathrm{L}(\tau_2,x)$, $\boldsymbol{\Sigma}^\mathrm{E}$ is a covariance matrix with elements ${\Sigma}^{E}_{ij}=\min(t_i,t_j)-t_it_j/t_f$,
\begin{equation}
    \boldsymbol{\mu}^\mathrm{E}_{-j|j}=\boldsymbol{\mu}^\mathrm{E}_{-il|j}=\begin{bmatrix}
        \mu^E_i \\ \mu^E_l
    \end{bmatrix} + \tfrac{b_j-\mu^E_j}{\Sigma^E_{jj}}\begin{bmatrix}
        \Sigma^E_{ij} \\ \Sigma^E_{lj}
    \end{bmatrix},
\end{equation}
and
\begingroup
\renewcommand*{\arraystretch}{1.4}
\begin{equation}
    \boldsymbol{\Sigma}^\mathrm{E}_{-j\mid j}\!\!=\!\boldsymbol{\Sigma}^\mathrm{E}_{il\mid j}\!=\!\! \begin{bmatrix} \Sigma^E_{ii}\!-\!(\Sigma^{E}_{ij})^2\!/\Sigma^{E}_{jj} & \!\Sigma^E_{il}\!-\!\Sigma^E_{ij}\Sigma^E_{lj}/\Sigma^E_{jj}\! \\ \!\Sigma^E_{il}\!-\!\Sigma^{E}_{ij}\Sigma^E_{lj}/\Sigma^E_{jj} &\!\! \Sigma^E_{ll}\!-\!(\Sigma^E_{lj})^2\!/\Sigma^E_{jj} \end{bmatrix}\!\!.\!
\end{equation}
\endgroup
with $i,j,l\in\{0,1,2\}$, and for $\boldsymbol{\mu}^\mathrm{E}_{-j|j}$ and $\boldsymbol{\Sigma}^\mathrm{E}_{-j\mid j}$, $i\neq j \neq l$.
\begin{rem}
    The expression in \eqref{eq:I2_res} represents the probability of the process remaining below the threshold until the last \gls{LoS} region, then exceeding the threshold in that region. Nothing constrains the process after $\tau_2$, so it includes cases where the process crosses the threshold and returns below it before $b_3$, and cases where it remains above. It is most significant when the system layout results in $g_\mathrm{L}(s,x)$ having its lowest values (i.e., $d(\Omega(s))$ is shortest) in that region.
\end{rem}
\begin{equation}
    I_3 = \int_0^{s_1}A(\tau_1)C(\tau_1)d\tau_1, \label{eq:I3_res}
\end{equation}
where 
\begin{multline}
    A(\tau_1)=N(g_\mathrm{L}(\tau_1,x);0,1+\tau_1) \\ \times \big[(\mu_A-g_\mathrm{L}'(\tau_1,x))\Phi(c_A)+\tfrac{\sigma_A}{\tau_1}\varphi(c_A)\big], \label{eq:Atau1}
\end{multline}
$\mu_A=\frac{g_\mathrm{L}(\tau_1,x)}{(1+\tau_1)}$, $\sigma_A=\sqrt{\frac{\tau_1}{1+\tau_1}}$, $c_A=\frac{b_0-\mu_A}{\sigma_A}$,
\begin{equation}
    \!\!C(\tau_1)\!=\!\Phi_3\!\big(\![b_1,\!b_2,\!b_3]; \![g_\mathrm{L\!}(\tau_1,x),\! g_\mathrm{L\!}(\tau_1,x),\! g_\mathrm{L\!}(\tau_1,x)], \!\boldsymbol{\Sigma}^\mathrm{C}\big)\!,\!\! \label{eq:Ctau1}
\end{equation}
and $\boldsymbol{\Sigma}^\mathrm{C}$ is a covariance matrix with elements $\Sigma^{C}_{ij}=\min(s_i,s_j)-\tau_1$ with $i,j\in\{1,2,3\}$.
\begin{rem}
 The expression in \eqref{eq:I3_res} represents the scenario of an early threshold crossing followed by a return below the threshold. This is an important term that $I_1$ misses due to it only checking boundary values. This term grows with $\sigma_\mathrm{SF}$, as larger fluctuations make an excursion followed by a reversion more likely, and with the length of region 1, as there are more locations for an excursion to occur.
\end{rem}
\begin{equation}
    I_4 = \int_0^{s_1}\int_{s_2}^{s_3}A(\tau_1)V(\tau_1,\tau_2)D(\tau_2)d\tau_2 d\tau_1, \label{eq:I4_res}
\end{equation}
where 
\begin{align}
    &V(\tau_1,\tau_2)\!=\!N(g_\mathrm{L}(\tau_2,x);g_\mathrm{L}(\tau_1,x),\tau_2\!-\!\tau_1\!)\!\big[\!\big(\!\tfrac{g_\mathrm{L}(\tau_2,x)-\mu^{V}_2}{\tau_2-s_2} \notag \\ &-\!g'_\mathrm{L}(\tau_2,x)\!\big)\Phi_2\!\big([b_1,b_2]; \boldsymbol{\mu}^\mathrm{V}\!\!,\boldsymbol{\Sigma}^\mathrm{V}\big)\!-\!\tfrac{\sigma_2^V}{\tau_2-s_2}M\!(c_1,c_2,\rho_V\!)\!\big]\!,\! \label{eq:Vtau1tau2}
\end{align}
\begin{equation}
    \!M\!(c_1,c_2,\rho_V\!){=}-\!\rho_V\varphi(c_1)\Phi\!\Big(\!\!\tfrac{c_2-\rho_Vc_1}{\sqrt{1-\rho_V^2}}\!\!\Big)\!- \varphi(c_2)\Phi\!\Big(\!\!\tfrac{c_1-\rho_Vc_2}{\sqrt{1-\rho_V^2}}\!\!\Big)\!,\!
\end{equation}
$\boldsymbol{\mu}^\mathrm{V}$ is a mean vector with elements $\mu^V_i=g_\mathrm{L}(\tau_1,x)+\frac{s_i-\tau_1}{\tau_2-\tau_1}(g_\mathrm{L}(\tau_2,x)-g_\mathrm{L}(\tau_1,x))$, $\boldsymbol{\Sigma}^\mathrm{V}$ is a covariance matrix with elements $\Sigma^V_{ij}=\min(s_i,s_j)-\tau_1-\frac{(s_i-\tau_1)(s_j-\tau_1)}{\tau_2-\tau_1}$, $\sigma_i^V=\sqrt{\Sigma^V_{ii}}$, $\rho_V=\frac{\Sigma_{12}^V}{\sigma_1^V\sigma_2^V}$ and $c_i=\frac{b_i-\mu^V_i}{\sigma_i^V}$, with $i\in\{1,2\}$.
\begin{rem}
    The expression in \eqref{eq:I4_res} is the probability that the process crosses the threshold in region 1, drops below again before the start of region 3, and crosses again in region 3. It is typically the smallest term due to it requiring two rare events to occur in the same realization.
\end{rem}

\section{Numerical Results}
\label{sec:numres}
Numerical results were generated to verify the analysis in Secs. \ref{sec:PAposerr}, \ref{sec:imperf_CSI} and \ref{sec:blockage} and investigate the impacts of deployment imperfections on an \gls{SU} \gls{PASS}. The attenuation factor along the dielectric is $\kappa=0.1$ dB/m and the effective refractive index of the dielectric is $\eta_\mathrm{eff}=1.4$ \cite{ding_flexible_2025}. We consider a path loss exponent of $\gamma=\gamma_\mathrm{L}=2.2$ for \gls{LoS} links and $\gamma_\mathrm{N}=4.5$ for \gls{NLoS} links. The carrier frequency is $f_\mathrm{c}=28$ GHz and the spatial decorrelation distances are $d_\mathrm{LS}=20$ m \cite{molisch_wireless_2011}, corresponding to $\alpha=0.05$, and $d_\mathrm{SS}=\lambda/2$. Where not otherwise specified, it is assumed that $L=1$ m, $\boldsymbol{\psi}_\mathbf{u}=(L/2,5)$ and $\sigma_\mathrm{SF}=8$ dB.

\subsection{Impact of Position Errors}

\begin{figure}[t]
    \centering
    \includegraphics[width=0.8\linewidth]{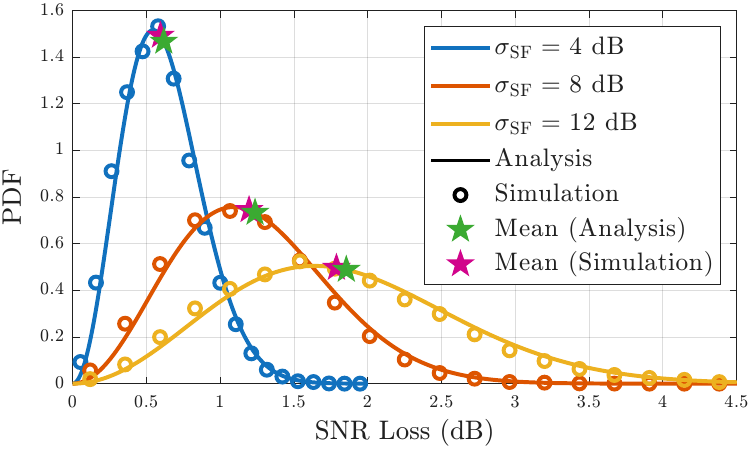}
    \caption{The PDF and mean of the SNR loss in dB of a \gls{PASS} with an $L=1$ m WG for a range of $\sigma_\mathrm{SF}$ values.}
    \label{fig:LossPDF}
\end{figure}

Figure \ref{fig:LossPDF} investigates the \gls{PDF} and mean of the \gls{SNR} loss for a range of $\sigma_\mathrm{SF}$ values when $L=1$ m and $d_\mathrm{err}=0.1$ m. Here, the expressions in Sec. \ref{sec:PAposerr} are converted to dB. 

Figure \ref{fig:LossPDF} shows that the derived expressions for both the \gls{PDF} and the mean closely match the simulation results for all considered values of $\sigma_\mathrm{SF}$. Increasing $\sigma_\mathrm{SF}$ increases the mean loss due to positioning errors, as stronger shadow fading fluctuations can produce larger channel-gain variations when the \gls{PA} is displaced from its optimal position. The distribution also broadens, indicating greater variability in the loss for a given positioning error. Thus, the mean alone does not fully capture the range of possible losses. Furthermore, the distributions exhibit slight positive skewness, with the mean lying to the right of the mode. This indicates that most losses occur around the peak value, while less frequent, larger losses shift the mean upwards.

\begin{figure}[t]
    \centering
    \includegraphics[width=0.8\linewidth]{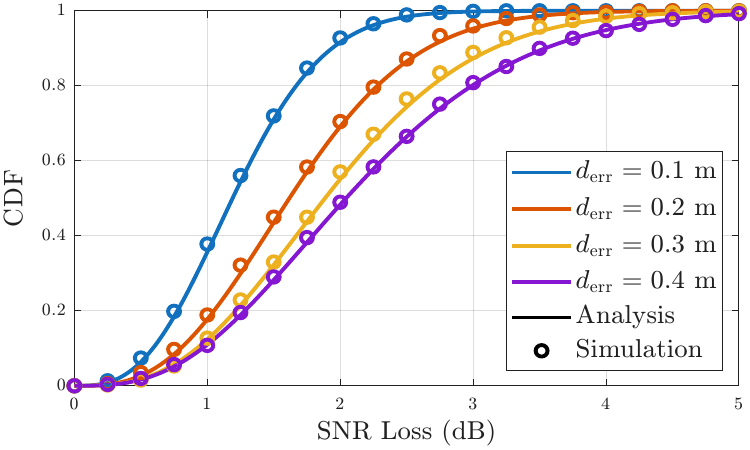}
    \caption{The CDF of the SNR loss in dB for a range of position errors.}
    \label{fig:LossCDF}
\end{figure}

Figure \ref{fig:LossCDF} compares the \gls{CDF} of the \gls{SNR} loss for position errors of $d_\mathrm{err}=\{0.1,0.2,0.3,0.4\}$ from the optimal position on an $L=1$ m \gls{WG}. The expressions in Sec. \ref{sec:PAposerr} are scaled such that the results are given in dB. For reference, for these parameters, the mean gain of placing the PA exactly at $x_\mathrm{p,LT}^\mathrm{opt}$ over random selection  is approximately 2 dB.

Figure \ref{fig:LossCDF} again shows excellent agreement between the analytical approximations and simulation. This is true even for large displacements, i.e., an error of 0.4 m corresponds to 40\% of the \gls{WG} length. This confirms that, over the considered range, the first-order Taylor approximation of the deterministic gain and the assumption of constant Brownian variance in the neighborhood of $x_\mathrm{p,LT}^\mathrm{opt}$ are not causing a significant loss of accuracy. As the displacement increases, the \glspl{CDF} shift to the right, indicating a higher probability of larger \gls{SNR} losses. A displacement of $d_\mathrm{err}=0.1$~m (10\% of the \gls{WG}) therefore removes approximately half of
the potential positioning gain and at $d_\mathrm{err}=0.4$~m, the median loss is comparable to the full gain.
Hence, although the loss in absolute terms is modest, the \gls{PA} must be placed
accurately for the benefit of \gls{LSF}-based position selection to be realized. The relative rightward shift diminishes as $d_\mathrm{err}$ grows, suggesting that the relationship between position error and \gls{SNR} loss becomes less sensitive as the antenna moves further from $x_\mathrm{p,LT}^\mathrm{opt}$.

\subsection{Impact of Imperfect CSI}

\begin{figure*}[t]
    \centering
     % First Figure
    \begin{minipage}{0.3\textwidth}
        \includegraphics[width=0.9\linewidth]{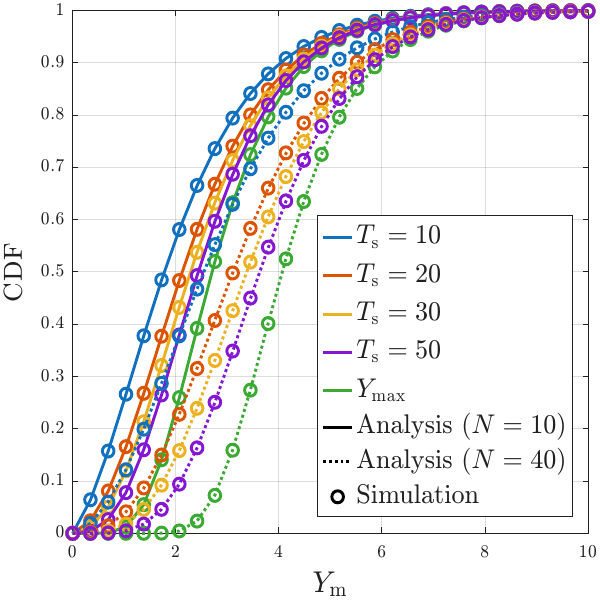}
        \caption{The CDF of $Y_\mathrm{m}$ corresponding to the best observed \gls{SSF} instance for a range of numbers of channel samples and measurements.}
        \label{fig:CSICDF}
    \end{minipage}
    \hfill % Adds horizontal space between minipages
    % Second Figure
    \begin{minipage}{0.3\textwidth}
        \centering
        \includegraphics[width=0.9\linewidth]{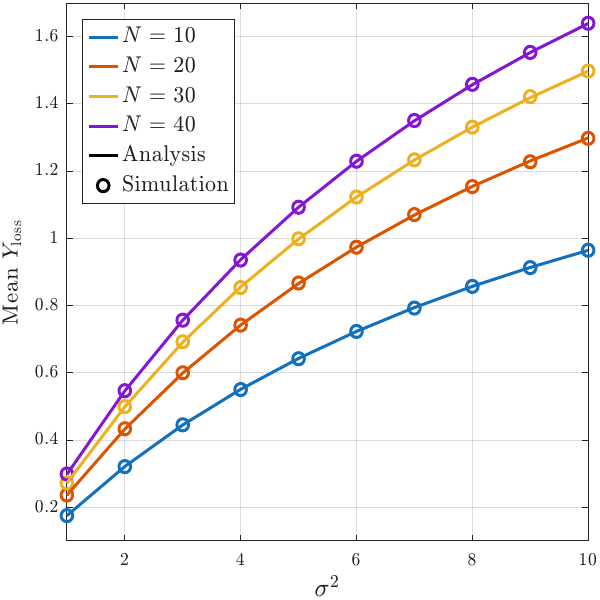}
        \caption{The mean loss due to imperfect CSI for a range of noise variances and \gls{SSF} sample numbers.}
        \label{fig:CSIloss}
    \end{minipage}
    \hfill
    % Third Figure
    \begin{minipage}{0.3\textwidth}
       \centering
        \includegraphics[width=0.9\linewidth]{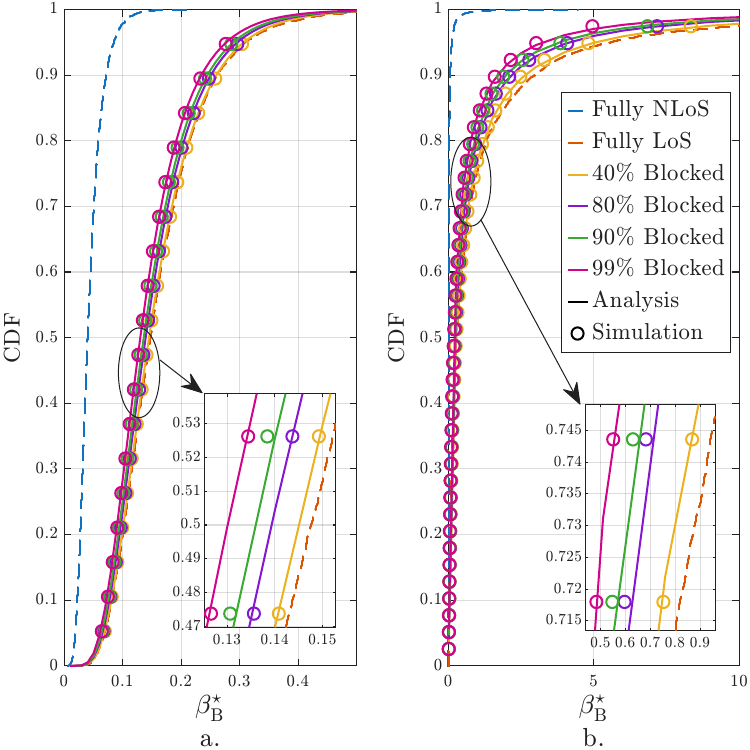}
        \caption{The CDF of $\beta_{\mathrm{B}}^\star$ for a range of blockage percentages across an $L=5$ m WG: a. $\sigma_\mathrm{SF}=2$~ dB, b. $\sigma_\mathrm{SF}=8$ dB.}
        \label{fig:CDFblockage}
    \end{minipage}
    \vspace{-1.25em}
\end{figure*}

Figure \ref{fig:CSICDF} investigates the distribution of the \gls{SCP} under imperfect \gls{CSI}. The \gls{CDF} of $Y_\mathrm{m}$ is plotted for multiple $T_\mathrm{s}$ values and compared with the \gls{CDF} of $Y_\mathrm{max}$. This is repeated for two values of $N$ to investigate how the number of independent \gls{SSF} samples affects the distribution. In all cases, $\sigma^2=8$, corresponding to a linear noise variance of approximately 9 dB relative to the unit channel power normalization.

Figure \ref{fig:CSICDF} shows excellent agreement between the analytical expressions in \eqref{eq:PYmlty} and \eqref{eq:PYmax} and simulations. As $T_\mathrm{s}$ increases, the \gls{CDF} of $Y_\mathrm{m}$ approaches that of $Y_\mathrm{max}$, since averaging more measurements reduces the estimation noise and thus improves the reliability of the \gls{PA} position selection. In the limit $T_\mathrm{s}\to\infty$, the channel estimates become exact and $Y_\mathrm{m}$ converges to $Y_\mathrm{max}$. Increasing the number of independent \gls{SSF} samples shifts the distributions towards higher channel powers, as a larger set of candidate locations increases the likelihood of finding a location with a strong channel realization. However, the benefit of increasing $N$ is less pronounced under imperfect \gls{CSI}. Here, selecting the location with the largest noisy estimate becomes more susceptible to estimation errors as the number of candidates grows. Thus, while increasing $N$ can improve the channel power, sufficiently accurate channel estimation is needed to realize this benefit.

Figure \ref{fig:CSIloss} studies the mean \gls{SCP} loss for a range of noise variances and \gls{SSF} sample numbers. For these results, $T_\mathrm{s}= 10$. Again, excellent agreement is shown between analysis and simulation. Combining \eqref{eq:EYloss}-\eqref{eq:EYm}, the mean loss simplifies to $\mathbb{E}[Y_{\mathrm{loss}}]
=(H_N-1)\frac{\sigma^2}{T_\mathrm{s}+\sigma^2}=(H_N-1)\frac{r}{r+1}$ where $r=\sigma^2/T_\mathrm{s}$. For all $N$, the mean loss increases with $\sigma^2$, as noisier estimates weaken the link between the measured and true \glspl{SCP}, making the selected
position less likely to be the optimal. The increase is concave rather than linear. The slope with respect to $\sigma^2$ is proportional to $T_s/(T_s+\sigma^2)^2$, so
the slope at $\sigma^2=10$ is approximately 30\% of that at $\sigma^2=1$. The
loss saturates at $H_N-1$, which is the loss incurred by random selection. This is
reached once the measured \gls{SCP} carries no information about the true \gls{SCP}.
Since selection cannot perform worse than random, further increases in noise
have a diminishing effect. Furthermore, as $\mathbb{E}[Y_\mathrm{loss}]$ depends on $\sigma^2$ and $T_s$ only through $r$, doubling the number of measurements
has the same effect as halving the noise variance. The mean loss also increases with $N$, consistent with Fig.~5. A larger
candidate set raises the achievable \gls{SCP} under perfect \gls{CSI} faster than
that obtained with noisy estimates, as more candidates provide more
opportunities for a noise-inflated position to be mistaken for the best. As $H_N-1$ grows logarithmically in $N$, the spacing between the curves shrinks as $N$ increases.

\subsection{Impact of LoS Blockages}
Figure \ref{fig:CDFblockage} investigates the impact of a single \gls{LoS} blockage on the \gls{CDF} of $\beta_\mathrm{B}^\star$ for a range of \gls{WG} blockage percentages. The blockage is centered in the middle of the $L=5$ m \gls{WG}. Fully blocked (i.e., fully \gls{NLoS}) and fully unblocked (i.e., fully \gls{LoS}) are shown as limiting cases. Results are shown for $\sigma_\mathrm{SF}=2$ dB (Fig. \ref{fig:CDFblockage}a) and $\sigma_\mathrm{SF}=8$ dB (Fig. \ref{fig:CDFblockage}b). 

The analytical expression in \eqref{eq:FBstar} accurately matches simulations.
In both Figs.~\ref{fig:CDFblockage}a and \ref{fig:CDFblockage}b, increasing the
blockage size shifts the \gls{CDF} to the left, as the position that would be
optimal without blockages becomes increasingly likely to be blocked. However,
the impact remains limited unless the entire \gls{WG} is blocked. Even with 99\%
of the \gls{WG} blocked, the \gls{CDF} of $\beta_\mathrm{B}^\star$ is very close to that of a fully \gls{LoS} system. This highlights that the creation of \gls{LoS} links is the most important role of a \gls{PASS}. Provided one \gls{LoS} position exists, performance near to that without blockages can be
obtained for the geometry considered. The loss grows with $\sigma_\mathrm{SF}$. With 99\% of the \gls{WG} blocked, the \gls{PA} can only select from a short \gls{LoS} segment over which the correlated shadowing varies very little, so the gain from searching for a favorable shadowing realization is lost. As this gain scales with $\sigma_\mathrm{SF}$, the resulting penalty is
larger for $\sigma_\mathrm{SF}=8$~dB than for
$\sigma_\mathrm{SF}=2$~dB. Additionally, increasing $\sigma_\mathrm{SF}$ extends the upper tail of $\beta_\mathrm{B}^\star$ relative to its median, as the best position benefits from more favorable realizations.

\section{Conclusion}

This paper has presented an analysis of the impact of three practical deployment imperfections on \gls{PASS} performance: \gls{PA} positioning errors, imperfect \gls{CSI}, and \gls{LoS} blockages. Analytical expressions were derived for the resulting SNR loss, scattered channel power, and optimum long-term \gls{SNR}, respectively, and validated through simulation. The results highlight distinct performance trade-offs. Positioning errors can reduce the gains from long-term position selection, with the loss increasing under stronger shadow fading. Imperfect \gls{CSI} introduces a selection penalty that depends on the ratio of noise variance to the number of channel measurements, while sufficiently favorable channel realizations remain achievable with the same upper-tail behavior as under perfect CSI. Most notably, \gls{LoS} blockages have a surprisingly limited impact: even when 99\% of the waveguide is blocked, performance remains close to that of a fully unblocked system for the considered geometry. This confirms that a key benefit of PASSs is their ability to create \gls{LoS} links through PA repositioning. These results demonstrate the potential of \glspl{PASS} to maintain robust performance under realistic deployment imperfections, while identifying the importance of accurate \gls{PA} positioning and channel estimation in realizing their full benefits.

\section*{Appendix A: Evaluating $G(z)$ and $D(T)$}
\noindent \textbf{Solving $G(z)$:}
Distributing the integral through the expression,
\begin{align}
    G(z) \!&=\!\!  \int_0^\infty\!\!\!\mathrm{e}^{-\frac{(w-z)^2}{2\upsilon}+\mu w}dw-\!\!\int_0^\infty\!\!\!\mathrm{e}^{-\frac{(w+z)^2}{2\upsilon}+\mu w} dw,\! \notag \\
    &= G_1 - G_2.
\end{align}
Completing the square and setting $x\!=\!(w\!-\!(z\!+\!\mu\upsilon))/\sqrt{\upsilon}$ gives
\begin{equation}
    G_1 = \sqrt{2\pi\upsilon}\mathrm{e}^{\mu z+\upsilon\mu^2/2}\Phi\left(\tfrac{z+\mu\upsilon}{\sqrt{\upsilon}}\right).
\end{equation}
Following the same process for $G_2$, and combining the results for $G_1$ and $G_2$ gives the final result for $G(z)$ in \eqref{eq:Gy}.

\noindent \textbf{Solving $D(T)$:}
Letting $w\!=\!\sqrt{T}z$ and $\theta_T\!=\!\mu \sqrt{T}$ and completing the square, 
\begin{equation}
    D(T)=T\mathrm{e}^{\theta_T^2/2}\int_0^\infty z\mathrm{e}^{-(z-\theta_T)^2/2}dz.
\end{equation}
Making another substitution, $x=z-\theta_T$, 
\begin{equation}
    D(T)\! = \!T\mathrm{e}^{\theta_T^2/2}\!\!\!\int_{-\theta_T}^\infty\!\!\! x\mathrm{e}^{-x^2\!/2}dx+\theta_T T \mathrm{e}^{\theta_T^2/2}\!\!\!\int_{-\theta_T}^\infty\!\!\!\mathrm{e}^{-x^2\!/2}dx.\!
\end{equation}
Integrating the first integral by parts, and recognizing the second as a normal \gls{CDF} gives \eqref{eq:Dt}.
\section*{Appendix B: The CDF and Mean of $\Lambda(u)$}
\noindent \textbf{Distribution:} Distributing the integral through the expression,
\begin{align}
    F_\Lambda(y) &= \!\tfrac{\sqrt{T}}{s\sqrt{s}}\tfrac{\mathrm{e}^{\upsilon \mu^2/2}}{\sigma_0(1\!+\!\theta_T\Phi(\theta_T)/\varphi(\theta_T))}\Big(\!\int_0^y\!\!\tfrac{z}{\sigma_0}\mathrm{e}^{\frac{z}{\sigma_0}(\mu-\frac{z}{2s\sigma_0})} \notag  \\ \times\Phi\Big(&\!\tfrac{z}{\sigma_0\sqrt{\upsilon}}\!+\!\theta_\upsilon\!\!\Big)dz\!-\!\!\!\int_0^y\!\!\tfrac{z}{\sigma_0}\mathrm{e}^{-\frac{z}{\sigma_0}(\mu+\frac{z}{2s\sigma_0})}\Phi\!\left(\!\theta_\upsilon\! -\! \tfrac{z}{\sigma_0\sqrt{\upsilon}}\!\right)\!dz\Big). \notag \\
    &=\!\tfrac{\sqrt{T}}{s\sqrt{s}}\tfrac{\mathrm{e}^{\upsilon \mu^2/2}}{\sigma_0(1\!+\!\theta_T\Phi(\theta_T)/\varphi(\theta_T))}(T_1-T_2). \label{eq:FLambday_T1T2}
\end{align}
To solve $T_1$, completing the square in the exponential and substituting $w=z/(\sigma_0\sqrt{s})-\theta_s$ leads to
\begin{align}
    T_1 =& \mathrm{e}^{\theta_s^2/2}s\sqrt{2\pi}\Big(\int^{y/(\sigma_0\sqrt{s})-\theta_s}_{-\theta_s}w\varphi(w)\Phi(a_\Lambda+b_\Lambda w)dw \notag \\ &+\theta_s \int^{y/(\sigma_0\sqrt{s})-\theta_s}_{-\theta_s}\varphi(w)\Phi(a_\Lambda+b_\Lambda w)dw\Big), \notag \\
    =&\mathrm{e}^{\theta_s^2/2}s\sqrt{2\pi}(T_{11} + T_{12}),
\end{align}
where $a_\Lambda{=} \theta_{T}/\sqrt{\upsilon}$ and $b_\Lambda{=}\sqrt{s/\upsilon}$. Using \cite[Eq. (10011.1)]{owen_table_1980} to solve $T_{11}$ and \cite[Eqs. (10010.1), (10010.4)]{owen_table_1980} to solve $T_{12}$,
\begin{align}
    &T_1 = \mathrm{e}^{\frac{\theta_s^2}{2}}\!s\sqrt{2\pi}\Big(\!\zeta_b\varphi(\zeta_a)\!\Big(\!\Phi\Big(\!\Theta_s(y)\sqrt{\!1\!+\!b_\Lambda^2}\!+\!a_\Lambda\zeta_b\!\Big)-\Phi\!\Big(\!\!-\!\theta_s\!\notag \\
    &\times\!\sqrt{\!1\!+\!b_\Lambda^2}\!+\!a_\Lambda\zeta_b\!\Big)\!\!\Big)\!\!-\!\Phi\big(\!a_\Lambda\!+\!b_\Lambda\Theta_{\!s\!}(y)\!\big)\varphi\big(\!\Theta_{\!s\!}(y)\!\big)\!\!+\!\varphi(\!-\theta_s\!)\Phi(a_\Lambda\notag \\ &-\!b_\Lambda\theta_s)\!+\!\theta_s\Big[\Phi_2\Big(\!\Theta_s(y);\zeta_a,-\zeta_b\Big) \!-\!\Phi_2\!\Big(\!\!-\!\theta_s;\zeta_a,\!-\zeta_b\!\Big)\!\Big]\!\Big)\!.\!\! \label{eq:T1}
\end{align}
with $\zeta_a=a_\Lambda/\sqrt{1+b_\Lambda^2}$, $\zeta_b=b_\Lambda/\sqrt{1+b_\Lambda^2}$ and $\Theta_s(y)=\tfrac{y}{\sigma_0\sqrt{s}}\!-\!\theta_s$.
To solve $T_2$, we can apply symmetry such that $\Phi(\theta_\upsilon-z/(\sigma_0\sqrt{\upsilon}))=1-\Phi(z/(\sigma_0\sqrt{\upsilon})-\theta_\upsilon)$, giving
\begin{align}
    T_2 \!&=\!\!\!\int_0^y\!\!\!\tfrac{z}{\sigma_0}\mathrm{e}^{\!-\frac{z}{\sigma_0}\!(\mu+\frac{z}{2s\sigma_0})}\!dz\!-\!\!\! \int_0^y\!\!\!\tfrac{z}{\sigma_0}\mathrm{e}^{\!-\frac{z}{\sigma_0}\!(\mu+\frac{z}{2s\sigma_0})}\Phi\big(\!\tfrac{z}{\sigma_0\sqrt{\upsilon}}\!-\!\theta_\upsilon\!\big)dz, \notag \\
    & = T_{21} - T_{22}.
\end{align}
Note that $T_{22}$ is equivalent to $T_1$ with $\mu$ replaced with $-\mu$. Therefore, only $T_{21}$ remains to be solved. Completing the square and making the substitution $w = z/(\sigma_0\sqrt{s})+\theta_s$,
\begin{multline}
    T_{21}=s\mathrm{e}^\frac{\theta_s^2}{2}\Big(\int_{\theta_s}^{y/(\sigma_0\sqrt{s})+\theta_s}w\mathrm{e}^{-w^2/2}dw \\ -\theta_s\int_{\theta_s}^{y/(\sigma_0\sqrt{s})+\theta_s}\mathrm{e}^{-w^2/2} dw \Big).
\end{multline}
Recognizing that the first term is the antiderivative of $-\mathrm{e}^{-w^2/2}$, and that the second is a normal probability,
\begin{equation}
    T_{21}\!=\!s-s\mathrm{e}^{-\frac{y}{\sigma_0}\!\left(\!\frac{y}{2s\sigma_0}+\mu\!\right)}\!\!-s\mathrm{e}^\frac{\theta_s^2}{2}\theta_s\!\sqrt{2\pi}\!\big(\!\Phi\!\big(\!\tfrac{y}{\sigma_0\sqrt{s}}\!+\!\theta_s\!\big)\!-\!\Phi(\theta_s)\!\big)\!. \notag
\end{equation}
Combining $T_{21}$ and $T_{22}$ gives $T_2$. Substituting $T_1$ in \eqref{eq:T1} and $T_2$ into \eqref{eq:FLambday_T1T2} gives the full distribution in \eqref{eq:FLambday}.

\noindent \textbf{Mean:} Distributing the integral through \eqref{eq:ELambdau_ints}, 
\begin{multline}
\mathbb{E}[\Lambda(u)]=\tfrac{\sqrt{T}}{s\sqrt{s}}\tfrac{\mathrm{e}^{\theta_\upsilon^2/2}}{(1+\theta_T\Phi(\theta_T)/\varphi(\theta_T))}\Big(\!\int_{0}^{\infty}\!\!y^2\mathrm{e}^{-\frac{y^2}{2s}+\mu y}  \\ \times\!\Phi\!\Big(\!\tfrac{y}{\sqrt{\upsilon}}\!+\!\theta_\upsilon\!\Big)dy - \!\!\int_0^\infty \!\!y^2\mathrm{e}^{-\frac{y^2}{2s}-\mu y}\Phi\!\Big(\!\theta_\upsilon\!-\!\tfrac{y}{\sqrt{\upsilon}}\!\Big)dy\!\Big), \notag \end{multline}
\begin{equation}
    \mathbb{E}[\Lambda(u)] = \tfrac{\sqrt{T}}{s\sqrt{s}}\tfrac{\mathrm{e}^{\theta_\upsilon^2/2}}{(1+\theta_T\Phi(\theta_T)/\varphi(\theta_T))}(P_1 - P_2).
\end{equation}
Completing the square in the exponential, making the substitution $w=y/\sqrt{s}-\theta_s$ and distributing the integrals,
\begin{align}
    P_1 &= \mathrm{e}^{\frac{\theta_s^2}{2}}s\sqrt{2\pi s}\Big[\!\int_{-\theta_s}^\infty\!\! w^2\varphi(w)\Phi(a_\Lambda\!+\! b_\Lambda w)dw \!+\! 2\theta_s\!\!\int_{-\theta_s}^\infty \!w \notag \\ &\times\!\varphi(w)\Phi(a_\Lambda\!+\!b_\Lambda w)dw \!+\! \theta_s^2\!\!\int_{-\theta_s}^\infty \varphi(w)\Phi(a_\Lambda \!+\! b_\Lambda w)dw\Big]\!, \notag \\ 
    &=\mathrm{e}^{{\theta_s^2}/{2}}s\sqrt{2\pi s}[P_{11} + 2\theta_sP_{12} + \theta_s^2P_{13}].
\end{align}
The integral in $P_{12}$ can be solved with \cite[Eq. (10011.1)]{owen_table_1980} and $P_{13}$ can be solved with \cite[Eq. (10010.1)]{owen_table_1980}. To solve $P_{11}$, let $z=-w$, and using the identities $\varphi'(z)=-z\varphi(z)$ and $\varphi''(z)=(z^2-1)\varphi(z)$,
\begin{align}
    P_{11} &= \!\!\int_{-\infty}^{\theta_s}\!\varphi''(z)\Phi(a_\Lambda\!-\! b_\Lambda z)dz\!+\!\!\int_{-\infty}^{\theta_s}\!\varphi(z)\Phi(a_\Lambda\!-\!b_\Lambda z)dz. \notag \\
    &= P_{111} + P_{112}.
\end{align}
Again, \cite[Eq. (10010.1)]{owen_table_1980} solves $P_{112}$. Using integration by parts, expanding the normal distribution in the resulting integral and completing the square
\begin{multline}
    P_{111}= -\theta_s\varphi(\theta_s)\Phi(a_\Lambda-b_\Lambda \theta_s)- b_\Lambda\varphi(\zeta_a) \\ \times\int_{-\infty}^{\theta_s}z\varphi\Big(z\sqrt{1+b_\Lambda^2}
    -a_\Lambda\zeta_b\Big)dz.
\end{multline}
Solving the integral via integration by parts, 
\begin{multline}
    P_{111} =
    -\theta_s\varphi(\theta_s)
    \Phi(a_\Lambda-b_\Lambda\theta_s)
    -\zeta_b\varphi(\zeta_a)\big(
    \tfrac{a_\Lambda b_\Lambda}{1+b_\Lambda^2}\\
    \times
    \Phi\big(\theta_s\sqrt{1\!+\!b_\Lambda^2}
    \!-\!a_\Lambda\zeta_b\big)
    \!-\!\tfrac{\zeta_b}{b_\Lambda}
    \varphi\big(\theta_s\sqrt{1\!+\!b_\Lambda^2}
    -a_\Lambda\zeta_b\big)\!\big).
    \label{eq:P112}
\end{multline}
Combining $P_{111}$ and $P_{112}$ to obtain $P_{11}$, and combining this with $P_{12}$ and $P_{13}$ gives $P_1$. For $P_2$, $\mu$ is  again replaced with $-\mu$, and using the symmetry of of the standard normal \gls{CDF},
\begin{align}
    P_2 &= \int_0^\infty y^2\mathrm{e}^{-\frac{y^2}{2s}-\mu y}dy- \int_0^\infty y^2\mathrm{e}^{-\frac{y^2}{2s}-\mu y}\Phi\Big(\tfrac{y}{\sqrt{\upsilon}}-\theta_\upsilon\Big)dy \notag \\
    & = P_{21} - P_{22}.
\end{align}
Note that $P_{22}$ is $P_1$ with $\mu$ replaced by $-\mu$. To solve $P_{21}$, recognize that it can be expressed as the second derivative of a simpler integral. Letting $P_{2d} \!=\!\! \int_0^\infty\exp(-\frac{y^2}{2s}\!-\!\mu y)dy$, $P_{21}\!=\! \frac{d^2P_{2d}}{d\mu^2}$. Completing the square and substituting $w\!=\!\frac{y}{\sqrt{2s}}\!+\!\frac{\theta_s}{\sqrt{2}}$,
\begin{equation}
    P_{2d}=\sqrt{2s}\mathrm{e}^\frac{\theta_s^2}{2}\!\!\int_{\theta_s/\sqrt{2}}^\infty\!\mathrm{e}^{-w^2}\!dw=\sqrt{\tfrac{\pi s}{2}}\mathrm{e}^\frac{\theta_s^2}{2}\mathrm{erfc}\left(\tfrac{\theta_s}{\sqrt{2}}\right)\!.
\end{equation}
To differentiate twice with respect to $\mu$, we use the fact that $\frac{d}{d\mu}\mathrm{erfc}(a\mu)=-2a/\sqrt{\pi}\exp(-a^2\mu^2)$, leading to
\begin{equation}
    P_{21} = \sqrt{2\pi}s^{3/2}(1+\theta_s^2)\mathrm{e}^{{\theta_s^2}/{2}}(1-\Phi(\theta_s))-s^2\mu.
\end{equation}
Therefore, combining $P_{21}$ and $P_{22}$ gives $P_2$, which, when combined with $P_1$ and simplified, gives $\mathbb{E}[\Lambda(u)]$ in \eqref{eq:ELambdau}.

\section*{Appendix C: The Moments and CDF of $Y_\mathrm{m}$}

\noindent \textbf{First Moment:}
As the measured channels are \gls{iid}, the corresponding true channels are also \gls{iid}, and thus the only dependence occurs between $(X_i,Y_i)$ pairs. If $R_1=\sqrt{X_i}$ and $R_2=\sqrt{Y_i}$, then the \gls{PDF} of $(R_1,R_2)$ is bivariate Rayleigh with the \gls{PDF} \cite[Eq. (3.9)]{simon_probability_2006}
\begin{multline}
    f_{R_1,R_2}(r_1,r_2) = \tfrac{r_1r_2}{\sigma_1^2\sigma_2^2(1-\rho^2)}\\ \times\exp\!\big(\!\tfrac{-1}{2(1-\rho^2)}\!\big(\!\tfrac{r_1^2}{\sigma_1^2}+\tfrac{r_2^2}{\sigma_2^2}\big)\!\big)  I_0\!\big(\!\tfrac{r_1r_2|\rho|}{\sigma_1\sigma_2(1-\rho^2)}\!\big)\!, \label{eq:fR1R2}
\end{multline}
where $\sigma_1^2=1/(2\lambda_T)$, $\sigma_2^2=1/2$ and 
\begin{align}
    \rho&=\mathrm{Corr}[\Re(h_i),\Re(\hat{h}_i)]=\mathrm{Corr}[\Re(h_i),\Re({h}_i)+\Re(z_i)], \notag \\
    &=\tfrac{1}{\sqrt{1+\sigma^2/T_\mathrm{s}}}=\sqrt{\lambda_T}.
\end{align}
Hence, the \gls{PDF} of $(X_i,Y_i)$, where $X_i=R_1^2$ and $Y_i=R_2^2$, can be obtained by scaling \eqref{eq:fR1R2} by the absolute value of the determinant of the Jacobian $J=1/(4\sqrt{xy})$, such that
\begin{align}
    f_{X,Y}(x,y) &= \tfrac{\exp\!\big(\!\frac{-1}{2(1-\rho^2)}\!\!\big(\!\frac{x}{\sigma_1^2}+\frac{y}{\sigma_2^2}\!\big)\!\big)}{4\sigma_1^2\sigma_2^2(1-\rho^2)}I_0\!\big(\!\tfrac{\sqrt{xy}|\rho|}{\sigma_1\sigma_2(1-\rho^2)}\!\big)\!, \notag \\
    &= a\exp\left(-(bx+cy)\right)I_0\left(d\sqrt{xy}\right), \label{eq:fXYxy}
\end{align}
where $a=\frac{1}{4\sigma_1^2\sigma_2^2(1-\rho^2)}=\frac{\lambda_T}{1-\lambda_T}$, $b=\frac{1}{2\sigma_1^2(1-\rho^2)}=\frac{\lambda_T}{1-\lambda_T}$, $c=\frac{1}{2\sigma_2^2(1-\rho^2)}=\frac{1}{1-\lambda_T}$, and $d=\frac{|\rho|}{\sigma_1\sigma_2(1-\rho^2)}=\frac{2\lambda_T}{1-\lambda_T}$. Next, we note that the \gls{CDF} and \gls{PDF} of $X_\mathrm{m}$ are
\begin{equation}
    P(X_\mathrm{m}\leq x)=F_{X}(x)^N=(1-\mathrm{e}^{-\lambda_T x})^N, 
\end{equation}
\begin{align}
    f_{X_{\mathrm{m}}}(x)&=N(1-\mathrm{e}^{-\lambda_Tx})^{N-1}\lambda_T\mathrm{e}^{-\lambda_Tx}, \notag \\
    &= N\lambda_T\sum_{j=0}^{N-1}\tbinom{N-1}{j}(-1)^j\mathrm{e}^{-\lambda_T(j+1)x}, \label{eq:fXmx}
\end{align}
and using \cite[Eq. (6.8.6)]{david_order_2003},
\begin{align}
    \mathbb{E}[Y_\mathrm{m}]&=\mathbb{E}[\mathbb{E}[Y\!\mid\!X\!=\!X_\mathrm{m}]] =\!\!\int_0^\infty\!\!\mathbb{E}[Y\!\mid\! X\!=\!x]f_{X_\mathrm{m}}(x)dx, \notag \\
    &=\ Na\sum_{j=0}^{N-1}\tbinom{N-1}{j}(-1)^j\!\int_0^\infty\!\int_0^\infty\mathrm{e}^{-(b+j\lambda_T)x} \notag \\&\qquad\qquad\qquad\qquad\times y\mathrm{e}^{-cy}I_0(d\sqrt{xy})dy\,dx,
\end{align}
where $f_X(x) = \lambda_T\mathrm{e}^{-\lambda_T x}$ as each $X_i$ follows an exponential distribution. Using \cite[Eq. (6.633.4)]{gradshteyn_table_2007} with a substitution of $x=t^2$ to solve the integral over $x$,
\begin{equation}
    \mathbb{E}[Y_\mathrm{m}]\!= \!Na\!\!\sum_{j=0}^{N-1}\!\!\tbinom{N-1}{j}\!\tfrac{(-1)^j}{b+j\lambda_T }\!\int_0^\infty \!\!\!\!y \mathrm{e}^{-(c-d^2\!/4(b+j\lambda_T\!))y}dy.\!\!
\end{equation}
We can use the same result to integrate over $y$ as long as $c>d^2/(4(b+j\lambda_T))$. This condition rearranges to $1>\lambda_T/(1+j(1-\lambda_T))$, which is always true as $0<\lambda_T < 1$. Thus, 
\begin{equation}
    \mathbb{E}[Y_\mathrm{m}]=Na\sum_{j=0}^{N-1}\tbinom{N-1}{j}\tfrac{(-1)^j}{(b+j\lambda_T)\left(c-\frac{d^2}{4(b+j\lambda_T)}\right)^{\!2}}.
\end{equation}
Substituting in the values for $a,b,c$ and $d$ and simplifying,
\begin{equation}
    \mathbb{E}[Y_\mathrm{m}] = N\sum_{j=0}^{N-1}\tbinom{N-1}{j}\tfrac{(-1)^j(1+j(1-\lambda_T))}{(j+1)^2}.
\end{equation}
Rewriting the numerator $1+j(1-\lambda_T)=(j+1)-j\lambda_T$, using the fact that $j/(j+1)^2=1/(j+1)-1/(j+1)^2$ and applying the standard binomial identities
\begin{equation}
    \sum_{j=0}^{N-1}\!\tbinom{N-1}{j}\!\tfrac{(-1)^j}{j+1}=\tfrac{1}{N}, \qquad \sum_{j=0}^{N-1}\!\tbinom{N-1}{j}\tfrac{(-1)^j}{(j+1)^2}=\tfrac{H_N}{N},\notag 
\end{equation}
where $H_N$ is the standard harmonic number, gives \eqref{eq:EYm}.

\noindent \textbf{Second Moment:} As $X_\mathrm{m}$ and $Y_\mathrm{m}$ are not independent, 
\begin{equation}
    \mathbb{E}\!\left[Y_\mathrm{m}^2\right]=\int_0^\infty\int_0^\infty y^2 f_{X_\mathrm{m},Y_\mathrm{m}}(x,y)dxdy.
\end{equation}
As all \gls{SSF} realizations are equally likely,
\begin{equation}
    \mathbb{E}[Y_\mathrm{m}^2]=\int_0^\infty\!\!\int_0^\infty Ny^2F_X(x)^{N-1}f_{X,Y}(x,y)dx\,dy,
\end{equation}
where $F_X(x)=1-\mathrm{e}^{-x\lambda_T}$. Using \eqref{eq:fXYxy}, expanding the order statistics term using the binomial theorem and using \cite{gradshteyn_table_2007}[Eq. (6.633.4)] to integrate over $x$ leads to  
\begin{equation}
    \mathbb{E}[Y_\mathrm{m}^2]\!=\!Na\!\!\sum_{j=0}^{N-1}\!\tbinom{N-1}{j}\!\tfrac{(-1)^j}{b+j\lambda_T}\!\int_0^\infty \!\!\!y^2\mathrm{e}^{-\left(\!c-\frac{d^2}{4(b+j\lambda_T)}\!\right)y}dy.\!
\end{equation}
Integrating by parts twice over $y$, initially with $u=y^2$, $dv=\mathrm{exp}(-(c-d^2/(4(b+j\lambda_T))))$ and then with $u=y$, $dv=-1/(c-d^2/(4(b+j\lambda_T)))\mathrm{exp}(-(c-d^2/(4(b+j\lambda_T))))$ gives 
\begin{equation}
    \mathbb{E}[Y_\mathrm{m}^2]=2Na\sum_{j=0}^{N-1}\!\tbinom{N-1}{j}\tfrac{(-1)^j}{(b+j\lambda_T)\!\big(c-\frac{d^2}{4(b+j\lambda_T)}\big)^{\!3}}.
\end{equation}
Following the same simplification steps as for the first moment leads to the expression in \eqref{eq:EYm2}.

\noindent \textbf{Distribution:}
The distribution of $Y_\mathrm{m}$ is conditional on the value of $X_\mathrm{m}$, and can be expressed as 
\begin{align}
    P(Y_\mathrm{m}<y)&=\mathbb{E}\left[P(Y_\mathrm{m}<y\mid X=X_\mathrm{m})\right] \notag, \\
    & =\int_{x=0}^\infty\int_{t=0}^y \tfrac{f_{Y,X}(t,x)}{f_X(x)}f_{X_\mathrm{m}}(x)dt\,dx.
\end{align}
Substituting in \eqref{eq:fXYxy} and \eqref{eq:fXmx}, and using \cite[Eq. (6.633.4)]{gradshteyn_table_2007} for the integral over $x$ followed by the trivial integral over $t$,
\begin{equation}
    P(Y_\mathrm{m}\!<y)\!=\!Na\!\sum_{j=0}^{N-1}\!\tbinom{N-1}{j}\!\tfrac{(-1)^j}{b+j\lambda_T}\!\! \bigg(\!\!\tfrac{1-\exp\big(\!-\big(c-\frac{d^2}{4(b+j\lambda_T)}\big)y\big)}{c-\frac{d^2}{4(b+j\lambda_T)}}\!\!\bigg)\!. \notag
\end{equation}
Substituting in the values for $a,b,c$ and $d$, and simplifying:
\begin{equation}
    \!P(Y_\mathrm{m}\!<y)\!=\!N\!\sum_{j=0}^{N-1}\!\!\tbinom{N-1}{j}\!\tfrac{(-1)^j}{j+1}\!\!\left(\!1\!-\!\exp\!\left(\!\!-\tfrac{(j+1)y}{1+j(1-\lambda_T)}\!\right)\!\!\right)\!\!.\!
\end{equation}
Finally, applying the binomial theorem such that $\sum_{j=0}^{N-1}\binom{N-1}{j}\frac{(-1)^j}{j+1}=\frac{1}{N}$ gives the result stated in \eqref{eq:PYmlty}.

\section*{Appendix D: Solving $I_2$, $I_3$ and $I_4$}

The derivation of $I_2$-$I_4$ is built upon the following identities:
\begin{itemize}
    \item From \cite[Eqs. (6.74)-(6.76)]{deisenroth_mathematics_2020}, the product of two Gaussian \glspl{PDF} leads to a new scaled Gaussian:
    \begin{multline}
        N(z;\mu_1,\sigma_1^2)N(z; \mu_2,\sigma_2^2)\\ =N(\mu_1;\mu_2,\sigma_1^2+\sigma_2^2)N(z;\mu_3,\sigma_3^2), \label{eq:T3}
\end{multline}
where $\mu_3=\sigma_3^2(\frac{\mu_1}{\sigma_1^2}+\frac{\mu_2}{\sigma_2^2})$ and $\frac{1}{\sigma_3^2}=\frac{1}{\sigma_1^2}+\frac{1}{\sigma_2^2}$.
\item For a random vector $\mathbf{X}=[X_1\dots X_n]^T$, applying the truncated first moment identity in \cite[Eq. (17)]{manjunath_moments_2021} with lower bounds of $-\infty$ and rearranging gives
\begin{multline}
    \int_{\mathbf{x}\leq \mathbf{b}} \!\!x_k N_n(\mathbf{x}; \boldsymbol{\mu}, \boldsymbol{\Sigma})d\mathbf{x} \!=\!\mu_k\Phi_n(\mathbf{b}; \boldsymbol{\mu}, \boldsymbol{\Sigma})\!-\!\!\sum_{j=1}^n\Sigma_{kj} \\ \times N(b_j; \mu_j, \Sigma_{jj})\Phi_{n-1}(\mathbf{b}_{-j}; \boldsymbol{\mu}_{-j\mid j}, \boldsymbol{\Sigma}_{-j\mid j}), \label{eq:T4}
\end{multline}
where $n$ is the number of dimensions, $\mathbf{b}=[b_1\dots,b_n]$, $\boldsymbol{\mu}=[\mu_1\dots\mu_n]^T$, $\boldsymbol{\Sigma}$ is a covariance matrix with elements $\Sigma_{ij}$, and $\Phi_{n-1}(\mathbf{b}_{-j}; \boldsymbol{\mu}_{-j\mid j}, \boldsymbol{\Sigma}_{-j\mid j})$ is the Gaussian \gls{CDF} of the remaining $n-1$ variables, conditioned on $X_j=b_j$. 
\item From \cite[Eqs. (2.81)-(2.82)]{bishop_pattern_2006}, if $(X,Y)$ are jointly Gaussian, then $X\mid Y=y$ is Gaussian with
\begin{align}
    \mu_{X|Y}&= \mu_X + \Sigma_{XY}\Sigma_{YY}^{-1}(y-\mu_Y) \label{eq:T61},\\
    \Sigma_{X\mid Y} & = \Sigma_{XX}-\Sigma_{XY}\Sigma_{YY}^{-1}\Sigma_{YX}.\label{eq:T62}
\end{align}
\end{itemize}
\noindent \textbf{Term 1:}
For simplicity, let $b_i=g_\mathrm{L}(s_i,x)$. As $w_3$ appears in one term that is solely dependent on it, the integrals over $w_i$ in $I_2$ can be split into the two components shown in \eqref{eq:I2_res}, where
\begin{equation}
    D(\tau_2)=\!\! \int_{-\infty}^{b_3}\!\!f_t(w_3\!\mid\! g_\mathrm{L}(\tau_2,x)\!)dw_3 \!=\!\Phi\!\left(\!\tfrac{b_3\!-\!g_\mathrm{L}(\tau_2,x)}{\sqrt{s_3\!-\!\tau_2}}\!\right)\!\!, \label{eq:Dtau2_og}
\end{equation}
\begin{multline}
    E(\tau_2) = \int_{-\infty}^{b_2}\int_{-\infty}^{b_1}\int_{-\infty}^{b_0}\varphi(w_0) f_t(w_1\mid w_0)f_t(w_2\mid w_1)  \\ \times f_\tau(\tau_2\mid w_2) dw_0dw_1dw_2.
\end{multline}
To solve $E(\tau_2)$, begin by substituting \eqref{eq:TA} for $f_\tau(\tau_2\mid w_2)$ and rewriting $N(g_\mathrm{L}(\tau_2,x);w_2,\tau_2-s_2)$ as $N(w_2; g_\mathrm{L}(\tau_2,x),\tau_2-s_2)$. This term represents the transition from $w_2$ at time $s_2$ to the final value $g_\mathrm{L}(\tau_2,x)$ at time $\tau_2$. A Gauss-Markov chain for $(w_0,w_1,w_2,w_f)$ is thus formed, where $w_f=g_\mathrm{L}(\tau_2,x)$. This can be factorized into the marginal density of $w_f$ and the conditional density of $(w_0,w_1,w_2)$ given $w_f$. We can then apply \eqref{eq:T61} and \eqref{eq:T62} with $X=(w_0,w_1,w_2)$, $Y=w_f$ and $y=g_\mathrm{L}(\tau_2,x)$. Letting $t_i=1+s_i$, as $t_i\leq t_f$, $\mathrm{Cov}(w_i,w_f)=t_i$, which leads to $\boldsymbol{\mu}^\mathrm{E}$ with elements $\mu^E_i=\frac{t_i}{t_f}g_\mathrm{L}(\tau_2,x)$ and $\boldsymbol{\Sigma}^\mathrm{E}$ with elements $\Sigma^E_{ij}=\min(t_i,t_j)-\frac{t_it_j}{t_f}$. Thus,
\begin{multline}
    E(\tau_2)\!=\!N(g_\mathrm{L}(\tau_2,x);0,1+\tau_2)\!\!\int_{D_2} \!\!N_3\!\left(\![w_0\,\, w_1\,\, w_2]; \boldsymbol{\mu}^\mathrm{E}\!, \boldsymbol{\Sigma}^\mathrm{E}\!\right)\\\times\left[\tfrac{g_\mathrm{L}(\tau_2,x)-w_2}{\tau_2-s_2}-g_\mathrm{L}'(\tau_2,x)\right]d\mathbf{w}.
\end{multline} 
Distributing the integrals through the expression,
\begin{align}
    E(\tau_2)\!&=\!N(g_\mathrm{L}(\tau_2,x);0,1+\tau_2)\bigg[\left(\tfrac{g_\mathrm{L}(\tau_2,x)}{\tau_2-s_2}-g_\mathrm{L}'(\tau_2,x)\right)\notag \\ &\quad\times\!\!\int_{D_2}  \!\!N_3\!\left(\![w_0\,\, w_1\,\, w_2]; \boldsymbol{\mu}^\mathrm{E}\!, \boldsymbol{\Sigma}^\mathrm{E}\!\right)d\mathbf{w}\!-\!\tfrac{1}{\tau_2-s_2}\!\int_{D_2}\!w_2 \notag \\ &\quad\times N_3\!\left(\![w_0\,\, w_1\,\, w_2]; \boldsymbol{\mu}^\mathrm{E}\!, \boldsymbol{\Sigma}^\mathrm{E}\!\right)d\mathbf{w}\bigg], \notag \\ 
    & = E_1 - E_2.
\end{align}
Recognizing that the integral in $E_1$ is a trivariate normal \gls{CDF} and applying \eqref{eq:T4} to $E_2$ gives the result in \eqref{eq:Etau2}.

\noindent \textbf{Term 2:} Since $w_0$ only appears in two terms that are solely dependent on it, the integrals over $w_i$ in $I_3$ can be split into the two components shown in \eqref{eq:I3_res}, where
\begin{equation}
        A(\tau_1) = \int_{-\infty}^{b_0} \!\varphi(w_0)f_\tau(\tau_1\mid w_0)dw_0, \label{Atau1_og}
\end{equation}
\begin{multline}
    C(\tau_1) \!=\!\int_{-\infty}^{b_1}\int_{-\infty}^{b_2}\int_{-\infty}^{b_3}\!f_t(w_1\mid g_\mathrm{L}(\tau_1,x))\\ \times f_t(w_2\mid w_1) f_t(w_3\mid w_2)dw_1dw_2dw_3.
\end{multline}
It can be observed that $C(\tau_1)$ is a trivariate normal \gls{CDF} of three Brownian increments. Since the process is conditioned to start from $g_\mathrm{L}(\tau_1,x)$ and subsequent Brownian increments have zero mean, the mean vector is $[g_\mathrm{L}(\tau_1,x)\,\,\,  g_\mathrm{L}(\tau_1,x)\,\,\, g_\mathrm{L}(\tau_1,x)]$. As the process starts at time $\tau_1$, the covariance between the process values at $s_i$ and $s_j$ is
$\Sigma^C_{ij}=\min(s_i,s_j)-\tau_1$. Thus, $C(\tau_1)$ takes the form seen in \eqref{eq:Ctau1}.

To solve $A(\tau_1)$, begin by substituting \eqref{eq:TA} for $f_\tau(\tau_1\mid~w_0)$. By symmetry, $N(g_\mathrm{L}(\tau_1,x);w_0,\tau_1) = N(w_0;g_\mathrm{L}(\tau_1,x),\tau_1)$. Thus, \eqref{eq:T3} can be applied to $N(w_0;g_\mathrm{L}(\tau_1,x),\tau_1)N(w_0;0,1)$, giving $N(g_\mathrm{L}(\tau_1,x);0,1+\tau_1)N(w_0;\mu_A,\sigma_A^2)$ with $\mu_A=\frac{g_\mathrm{L}(\tau_1,x)}{1+\tau_1}$ and $\sigma_A^2=\frac{\tau_1}{\tau_1+1}$. Distributing the integral,
\begin{align}
    A(&\tau_1)\!=\!N(g_\mathrm{L}(\tau_1,x);0,1\!+\!\tau_1)\Big(\Big[\!\tfrac{g_\mathrm{L}(\tau_1,x)}{\tau_1}\!-\!g_\mathrm{L}'(\tau_1,x)\Big] \notag \\ &\!\!\times\!\! \int_{-\infty}^{b_0}\!\!N(w_0;\mu_A,\sigma_A^2)dw_0\!-\!\tfrac{1}{\tau_1}\!\int_{-\infty}^{b_0}\!\!w_0N(w_0;\mu_A,\sigma_A^2)dw_0\!\!\Big)\!\notag, \\
     &\quad\,\,= A_1 - A_2.
\end{align}
Recognizing that the integral in $A_1$ is a normal \gls{CDF}, and applying \eqref{eq:T4} to solve $A_2$ gives the final result seen in \eqref{eq:Atau1}.

\noindent \textbf{Term 3:}
As $w_0$ and $w_3$ each appear in one term that is solely dependent on it, the integrals over $w_i$ in $I_4$ can be split into the three components shown in \eqref{eq:I4_res}, where $A(\tau_1)$ is stated in \eqref{Atau1_og} and solved in \eqref{eq:Atau1}, $D(\tau_2)$ is given in \eqref{eq:Dtau2_og}, and  
\begin{multline}
    V(\tau_1,\tau_2)=\! \int_{-\infty}^{b_2}\int_{-\infty}^{b_1} f_t(w_1\mid g_\mathrm{L}(\tau_1,x))\\\times f_t(w_2\mid w_1)f_\tau(\tau_2\mid w_2)dw_1dw_2. 
\end{multline}
We begin similarly to $E(\tau_2)$ by substituting \eqref{eq:TA} for $f_\tau(\tau_2\mid~w_2)$, rewriting $N(g_\mathrm{L}(\tau_2,x);w_2,\tau_2-s_2)$ as $N(w_2; g_\mathrm{L}(\tau_2,x),\tau_2-s_2)$, and recognizing the resulting Gauss-Markov chain, this time for $(w_1,w_2,w_f)$. Again, we can factorize this into the marginal density of $w_f$ and the conditional density of $(w_1,w_2)$ given $w_f$, and thus let $X=(w_1,w_2)$ and $Y=w_f$. However, a key difference is that the process is conditioned on the known value $g_\mathrm{L}(\tau_1,x)$ at time $\tau_1$, so the chain evolves from this fixed starting point. Applying \eqref{eq:T61} and \eqref{eq:T62}, this time with $\boldsymbol{\mu}_X=[g_\mathrm{L}(\tau_1,x)\,\,g_\mathrm{L}(\tau_1,x)]^T$, $\mu_Y=g_\mathrm{L}(\tau_1,x)$, $y=g_\mathrm{L}(\tau_2,x)$, $\mathrm{Cov}(w_i,w_j)=\min(s_i,s_j)-\tau_1$, $\mathrm{Cov}(w_i,w_f)=s_i-\tau_1$ and $\mathrm{Cov}(w_f,w_f)=\tau_2-\tau_1$, we obtain $\boldsymbol{\mu}^\mathrm{V}$ with elements $\mu^\mathrm{V}_i=g_\mathrm{L}(\tau_1,x)+\frac{s_i-\tau_1}{\tau_2-\tau_1}(g_\mathrm{L}(\tau_2,x)-g_\mathrm{L}(\tau_1,x))$ and $\boldsymbol{\Sigma}^\mathrm{V}$ with elements $\Sigma^\mathrm{V}_{ij}=\min(s_i,s_j)\!-\!\tau_1\!-\!\frac{(s_i-\tau_1)(s_j-\tau_1)}{\tau_2-\tau_1}$. Therefore,
\begin{multline}
    V\!(\tau_1,\tau_2)\!=\!N(g_\mathrm{L}(\tau_2,x);g_\mathrm{L}(\tau_1,x),\tau_2-\tau_1)\! \!\int_{-\infty}^{b_2}\!\!N_2\Big(\![w_1\, w_2]; \\ \boldsymbol{\mu}^\mathrm{V}\!,\boldsymbol{\Sigma}^\mathrm{V}\Big)\!\left[\tfrac{g_\mathrm{L}(\tau_2,x)-w_2}{\tau_2-s_2}-\!g_\mathrm{L}'(\tau_2,x)\right]dw_1dw_2.
\end{multline}
Distributing the integrals through the expression,
\begin{align}
    V\!(\tau_1,\!\tau_2)\!&=\!N\!(g_\mathrm{L}(\tau_2,x),g_\mathrm{L}(\tau_1,x),\!\tau_2\!-\!\tau_1\!)\!\Big[\!\Big(\!\tfrac{g_\mathrm{L}(\tau_2,x)}{\tau_2-s_2}\!-\!g_\mathrm{L}'\!(\tau_2,x)\!\Big)\notag\\ &\times\!\!\int_{-\infty}^{b_2}\int_{-\infty}^{b_1}N_2\!\left(\![w_1\,w_2];\boldsymbol{\mu}^\mathrm{V}\!\!,\boldsymbol{\Sigma}^\mathrm{V}\right)\!dw_1dw_2\!-\!\tfrac{1}{\tau_2-s_2} \notag \\ &\times\!\!\int_{-\infty}^{b_2}\int_{-\infty}^{b_1}w_2N_2\!\left(\![w_1\,w_2];\boldsymbol{\mu}^\mathrm{V}\!\!,\boldsymbol{\Sigma}^\mathrm{V}\!\right)\!dw_1dw_2\Big]\!. \notag \\
    &= V_1 - V_2.
\end{align}
Again, recognizing that the integral in $V_1$ is a bivariate normal \gls{CDF} and applying \eqref{eq:T4} to $V_2$ gives \eqref{eq:Vtau1tau2}.

\bibliographystyle{IEEEtran}
\bibliography{references}

\end{document}